\documentclass[]{aastex}

\usepackage{emulateapj5}
\usepackage{onecolfloat}
\usepackage{graphicx} 
\usepackage{fancyheadings} 
\usepackage{ulem}
\usepackage{rotating}
\usepackage{lscape}

\def\intl{\int\limits}

\def\perd{\;\;\; .}
\def\cmma{\;\;\; ,}

\def\ltp{\left ( \,}

\def\rtp{\, \right  ) }

\def\smm{\sum\limits}
\def\avgq#1{<#1>}
\def\ohf{\frac{1}{2}}
\newcommand{\snrlim}{SNR$_{lim}$}
\newcommand{\nhi}{$N_{\rm HI}$}
\newcommand{\mnhi}{N_{\rm HI}}
\newcommand{\flls}{f_{\rm HI}^{\rm LLS}}
\newcommand{\fdla}{f_{\rm HI}^{\rm DLA}}
\newcommand{\llls}{$\ell_{\rm LLS}$}
\newcommand{\ldla}{\ell_{\rm DLA}}
\newcommand{\fnhi}{$f_{\rm HI}(N,X)$}
\newcommand{\mfnhi}{f_{\rm HI}(N,X)}
\newcommand{\Nth}{2 \sci{20} \cm{-2}}

\newcommand{\gz}{$g(z)$}
\newcommand{\omg}{$\Omega_g$}
\newcommand{\ostr}{$\Omega_*$}
\newcommand{\momg}{\Omega_g}
\newcommand{\olls}{$\Omega_g^{\rm LLS}$}
\newcommand{\odla}{$\Omega_g^{\rm DLA}$}
\newcommand{\oneut}{$\Omega_g^{\rm Neut}$}
\newcommand{\ohi}{$\Omega_g^{\rm HI}$}
\newcommand{\olwz}{$\Omega_g^{\rm 21cm}$}

\newcommand{\kms}{km~s$^{-1}$ }
\newcommand{\cm}[1]{\, {\rm cm^{#1}}}
\newcommand{\mkms}{{\rm \; km\;s^{-1}}}

\newcommand{\lya}{Ly$\alpha$}

\newcommand{\N}[1]{{N({\rm #1})}}
\newcommand{\sci}[1]{{\rm \; \times \; 10^{#1}}}

\begin{document}

\twocolumn[%
\submitted{Accepted to ApJ: August 15, 2005}
\title{The SDSS Damped \lya\ Survey: Data Release 3}

\author{Jason X. Prochaska \& St\'ephane Herbert-Fort}
\affil{Department of Astronomy and Astrophysics, 
UCO/Lick Observatory;
University of California, 1156 High Street, Santa Cruz, CA  95064;
xavier@ucolick.org, shf@ucolick.org}
\and
\author{Arthur M. Wolfe}
\affil{Department of Physics, and 
Center for Astrophysics and Space Sciences, 
University of California, San
Diego, 
Gilman Dr., La Jolla; CA 92093-0424; awolfe@ucsd.edu}

\begin{abstract}
We present the results from a damped \lya\ survey of the Sloan
Digital Sky Survey, Data Release 3.  We have discovered over 500
new damped \lya\ systems at $z>2.2$ and the complete 
statistical sample for $z>1.6$ has more than 600 damped \lya\ galaxies.
We measure the \ion{H}{1} column density distribution \fnhi\
and its zeroth and first moments 
(the incidence $\ldla$ and gas mass-density \odla\
of damped \lya\ systems, respectively)
as a function of redshift.
The key results include: 
(1) the full SDSS-DR3 \fnhi\ distribution 
($z \sim 3.06$) is well fit by
a $\Gamma$-function (or double power-law) with `break' column density 
$N_\gamma = 10^{21.5 \pm 0.1} \cm{-2}$ and `faint-end' slope
$\alpha = -1.8 \pm 0.1$;
(2) the shape of the \fnhi\ distributions in a series of redshift bins 
does not show evolution; 
(3) the incidence and gas mass density of damped systems
decrease by $35 \pm 9\%$ and $50 \pm 10\%$ during $\approx 1$\,Gyr
between the redshift interval $z=\lbrack 3.,3.5 \rbrack$ 
to $z=\lbrack 2.2,2.5 \rbrack$;
and (4) the incidence and gas mass density of damped \lya\
systems in the lowest SDSS redshift bin ($z=2.2$) are 
consistent with the current values.
We investigate a number of systematic errors in damped \lya\
analysis and identify only one important
effect: we measure $40 \pm 20\%$ higher \odla\
values toward a subset of brighter quasars than toward a faint subset.  
This effect is contrary to the bias associated with
dust obscuration and suggests that 
gravitational lensing may be important.
Comparing the results against several models of galaxy formation in 
$\Lambda$CDM, we find all of the models significantly underpredict 
$\ldla$ at $z=3$ and only SPH models 
with significant feedback (Nagamine et al.)
may reproduce \odla\ at high redshift.
Based on our results for the damped \lya\ systems, we
argue that the Lyman limit systems contribute 
$\approx 33\%$ of the universe's
\ion{H}{1} atoms at all redshifts $z=2$ to 5.
Furthermore, we infer that the \fnhi\ distribution 
for $\mnhi < 10^{20} \cm{-2}$ has an inflection with slope 
$d \log f/d \log N > -1$.
We advocate a new mass density definition -- the mass density
of predominantly 
neutral gas \oneut\ -- to be contrasted with the mass density of gas
associated with \ion{H}{1} atoms.
We contend the damped \lya\ systems contribute
$>80\%$ of \oneut\ at all redshifts and therefore are the main
reservoirs for star formation.


\end{abstract}
]

\pagestyle{fancyplain}
\lhead[\fancyplain{}{\thepage}]{\fancyplain{}{PROCHASKA, HERBERT-FORT, \& WOLFE}}
\rhead[\fancyplain{}{The SDSS Damped \lya\ Survey: DR3 }]{\fancyplain{}{\thepage}}
\setlength{\headrulewidth=0pt}
\cfoot{}

\section{Introduction}

The damped \lya\ systems are the class of quasar absorption 
line systems with \ion{H}{1} column density $\mnhi \geq \Nth$
\citep[see][for a review]{wolfe05}.  Unlike the \lya\ forest,
the damped \lya\ systems are comprised of predominantly neutral gas
and are proposed to be the progenitors of galaxies like the Milky Way
\citep[e.g.][]{kauff96}.
\cite{wolfe86} established the 
$N_{th} = \Nth$ threshold primarily to correspond
to the surface density limit of local 21\,cm observations at that time. 
It is somewhat fortuitous that this threshold roughly corresponds to the
transition from primarily ionized gas to predominantly neutral
gas \citep[e.g.][]{viegas95,pro96,pro99,vladilo01}.  

For the past two decades, several groups have surveyed high $z$
quasars for the damped \lya\ systems 
\citep{wolfe86,wolfe95,storrie96,storrie00,peroux03}.  These surveys
measured the \ion{H}{1} frequency distribution function and its
moments: the
incidence of the damped \lya\ systems $\ldla$ and the gas mass density
of these galaxies \odla.  
The latter quantity has cosmological significance.
Its evolution constrains the build-up
of structure within hierarchical cosmology \citep[e.g.][]{ma94,kly95},
it serves as the important neutral gas reservoir for star formation
at high redshift and
describes the competition between gas accretion and star formation
\citep[e.g.][]{fall93}, and it constrains models of galaxy
formation in hierarchical cosmology
\citep[e.g.][]{spf01,cen03,nagamine04a}.
Previous surveys have reported statistical error on \odla\ of
$\approx 30\%$ in redshift intervals $\Delta z \approx 0.5$ at high redshift.
As we enter the so-called era of precision cosmology, we aspire to
constrain \odla\ to better than
10$\%$.  Although not formally a cosmological parameter, a precise
determination of \odla\ and its redshift evolution
are fundamental constraints on any cosmological theory of galaxy formation.

\begin{table*}
\begin{center}
\caption{SDSS-DR3 QUASAR SAMPLE\label{tab:qso}}
\begin{tabular}{ccclclllll}
\tableline
\tableline
Plate &MJD & FiberID &Name &$z_{qso}$ & $f_{BAL}^a$ & $z_{i}$ &
$z_{i}^b$ & $z_{f}$ &$z_{candidate}$ \\
\tableline
 650&52143& 178&J000050.60$-10$2155.8&2.640&0&2.200&2.200&2.604&\\
 750&52235& 550&J000143.41$+15$2021.4&2.638&0&2.200&2.200&2.602&\\
 650&52143& 519&J000159.12$-09$4712.4&2.308&0&2.203&2.250&2.275&\\
 387&51791& 556&J000221.11$+00$2149.4&3.057&0&2.200&2.209&3.016&1.958\\
 650&52143& 111&J000238.41$-10$1149.8&3.941&0&3.203& ... & 3.891&3.271,3.523,3.605,3.655\\
 750&52235& 608&J000300.34$+16$0027.7&3.675&0&3.285&3.480&3.629&\\
 650&52143&  48&J000303.34$-10$5150.6&3.647&2&... & ... &2.897,3.465\\
 750&52235&  36&J000335.21$+14$4743.6&3.484&0&... & ... &\\
 650&52143& 604&J000413.63$-08$5529.5&2.424&0&2.200&2.200&2.389&\\
 650&52143& 605&J000424.16$-08$5047.9&2.433&0&2.200&2.200&2.399&\\
\tableline
\end{tabular}
\end{center}
\tablenotetext{a}{0=No intrinsic absorption; 1=Mild intrinsic absorption, included in analysis with restriction; 2=Strong intrinsic absorption,  excluded.}
\tablenotetext{b}{Starting redshift for SNR$_{lim} = 5$.}
\tablecomments{[The complete version of this table is in the electronic edition of the Journal.  The printed edition contains only a sample.]}
\end{table*}

In \cite{ph04}, hereafter PH04, we initiated a survey for
the damped \lya\ systems in the quasar spectra of the 
Sloan Digital Sky Survey (SDSS) Data Release 1 (DR1).  We
demonstrated that the spectral resolution, signal-to-noise
ratio (SNR), and wavelength coverage
of the SDSS spectra are well suited to survey the damped 
\lya\ systems at $z>2.2$.   We reported on the number of damped \lya\
systems
per unit redshift and the neutral gas mass density.
In this paper, we extend the survey to
include the full Data Release~3.  In
addition to substantially increasing the PH04 sample, this paper 
extends the analysis to include a determination of the 
$\N{HI}$ frequency distribution, \fnhi.  Furthermore, we perform 
a series of tests to examine systematic error related to the
analysis.  With the increased sample size, we believe the systematic
uncertainty is as important as statistical uncertainty.
Similarly, uncertainty related to selection bias (e.g.\ dust obscuration)
must be given careful attention.
Future progress will require significant strides on each of these three
fronts.

This paper is organized as follows: $\S$~\ref{sec:redpath}
describes the SDSS quasar sample and defines the sub-set used to
survey the damped \lya\ systems;
we present the damped \lya\ candidates and the \nhi\ measurements
in $\S$~\ref{sec:dla}, we
perform standard analysis of the \ion{H}{1} distribution in $\S$~\ref{sec:HI},
and we discuss the results in $\S$~\ref{sec:discuss}.
Aside from comparisons against local observations, we mainly restrict the
analysis to optical surveys for the damped \lya\ systems (i.e.\ $z>1.6$).  
Unless it is otherwise stated all log values and expressions
correspond to log base~10.
Here and throughout the paper we adopt 
cosmological parameters consistent with the latest WMAP results
\citep{bennett03}: 
$\Omega_\Lambda = 0.7, \Omega_m=0.3, H_0=70$\kms\,Mpc$^{-1}$.

\section{SDSS SAMPLE}
\label{sec:redpath}

\subsection{Redshift Path}

The quasar sample considered in this paper includes 
every object identified spectroscopically as a quasar
with $z>2.2$ in the SDSS-DR3\footnote{Specifically, where
SPECCLASS=3 or 4 in SPECPHOTO for SDSS-DR3.}.
We also include the full
compilation of quasars and damped \lya\ systems 
(avoiding duplication) from
the previous two decades of research as compiled by
\cite{peroux03}.
In each of these SDSS quasars we have defined a redshift
interval where we search for the presence of damped \lya\
systems.  In this paper, we define the redshift path using
an algorithm similar to that introduced in PH04.  
We also apply more conservative criteria to investigate the effects
of SNR (the estimated flux divided by estimated noise).

The starting redshift $z_i$ is defined as follows.
First, we identify the minimum wavelength $\lambda_i$
where the median SNR in a 20 pixel interval
exceeds \snrlim.   
In PH04, we took \snrlim~$= 4$ and in this paper we also consider
larger values.
An assumption of our prescription is that the median SNR does not 
decrease significantly at wavelengths greater than $\lambda_i$
unless a damped \lya\ candidate is present.  
This is a good assumption unless $\lambda_i$ coincides with the
Ly$\beta$/\ion{O}{6} 
emission feature of the quasar.  
At these wavelengths, the SNR of the data is elevated and it is possible
the median SNR may be above \snrlim\ in this spectral region but below
\snrlim\ otherwise.  Therefore, in the 10000\kms\ region centered at
$\lambda = 1025.7 (1+z_{qso})$ we demand that the median SNR be greater
than 3*\snrlim\ if $\lambda_i$ is to be set within that region.
In practice, this removes a number of faint quasars from the sample whose
spectra have median SNR greater than \snrlim\ only in the Ly$\beta$/\ion{O}{6}
emission line.
Finally, we define the starting redshift for the damped \lya\ search:
\begin{equation}
z_i \equiv (\lambda_i/1215.67) - 1. + 0.005  \cmma
\end{equation}
\noindent where the 0.005 increment offsets $z_i$ by 1500\,\kms.
Finally, we demand $z_i \geq 2.2$. 

Similarly, we define an ending redshift for the statistical
pathlength of a given quasar,
\begin{equation}
z_f \equiv 0.99 z_{qso} - 0.01 \cmma
\end{equation}
\noindent 
which corresponds to 3000\,\kms blueward of the \lya\ emission feature.
The offset minimizes the likelihood that a damped \lya\ system is physically
associated with the quasar.  For reasons unknown to us, a small fraction
of the quasar spectra have extended regions ($>50$ pixels)
with zero flux and null
values in their error array.  In these cases, we set $z_f$ to 
1000\,\kms blueward of the `null region'.

Each quasar spectrum was visually inspected and characterized
according to the presence or absence of features identified with
intrinsic absorption (e.g.\ BAL quasars).  
This is a necessary step in damped \lya\ surveys because
quasars with strong
intrinsic \ion{N}{5} and \ion{O}{6} absorption can be confused
with the \lya\ transition of an intervening quasar absorption
line system.  
Following PH04, we divided the quasars into three
categories: (a) quasars without significant intrinsic absorption;
(b) mild absorption-line quasars which show modest absorption at
the \ion{C}{4} emission feature; and (c) strong absorption-line
quasars whose \ion{C}{4}, \ion{O}{6} lines have large equivalent
widths and could be confused with a damped \lya\ system.
The latter category is discarded from all subsequent analysis.
For the mild absorption-line quasars, however, we search for
damped \lya\ systems in the redshift interval
$max(z_i,z_{BAL}) < z < min(z_{qso} - 0.08226,z_f)$ where 
$z_{BAL} \equiv (1+z_{qso}) (1060/\lambda_{Ly\alpha}) - 1$ and
the modification to the ending redshift
minimizes the likelihood of misidentifying
\ion{N}{5} absorption as a damped \lya\ system.

Table~\ref{tab:qso} presents the full list of SDSS-DR3 quasars
analyzed here.  Columns~1-9 designate the SDSS 
plate, MJD, and fiber numbers, the quasar name, 
the emission redshift, a flag describing intrinsic absorption,
$z_i$ and $z_f$ for \snrlim=4, and the absorption redshift of any damped \lya\
candidates along the sightline.  The latter may include candidates
where $z_{abs}$ is not in the $[z_i,z_f]$ interval.
Also, many candidates are false positive detections, in particular
BAL absorption lines.

\begin{figure}[ht]
\begin{center}
\includegraphics[height=3.6in,angle=90]{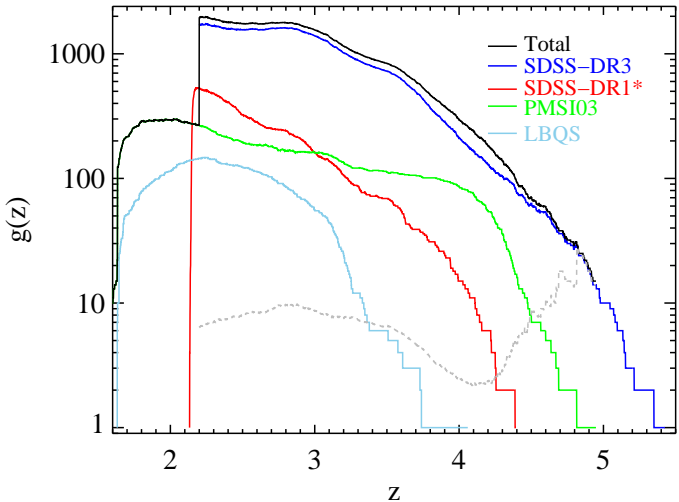}
\caption{Redshift sensitivity function $g(z)$ as a function of redshift for
the LBQS survey, the compilation of \cite{peroux03}, the SDSS-DR1
sample from PH04, the SDSS-DR3 sample studied here, and the total
sample.   The gray dashed-line traces the ratio of the SDSS-DR3 sample
to the previous 20 year compilation \citep{peroux03}.
}
\label{fig:gzsmpl}
\end{center}
\end{figure}

\subsection{\gz}

\begin{figure}[ht]
\begin{center}
\includegraphics[height=3.6in,angle=90]{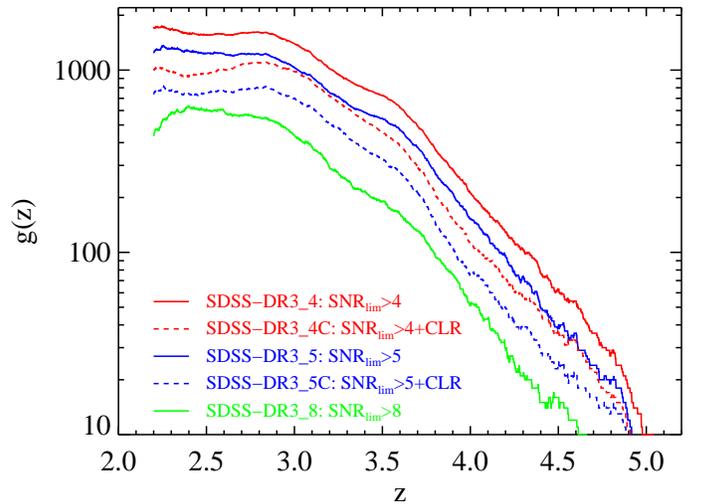}
\caption{Redshift sensitivity function $g(z)$ as a function of redshift for
several cuts of the SDSS-DR3 data (Table~\ref{tab:cuts}).
}
\label{fig:gzsdss}
\end{center}
\end{figure}

The survey size of a quasar absorption line study is 
characterized through the redshift sensitivity function \gz.
Figure~\ref{fig:gzsmpl} presents a series of \gz\ curves for the 
the Large Bright Quasar Survey (LBQS) search for damped \lya\
systems \citep{wolfe95}, the compilation of \cite{peroux03}, and 
the SDSS-DR1 sub-sample of PH04.  These are compared against
the full SDSS-DR3 sample with \snrlim~=4.
Note that the SDSS data releases are inclusive and also that the
\cite{peroux03} compilation includes the LBQS sample.
Comparing the curves, we find
the current SDSS sample now 
exceeds the previous surveys by an order of magnitude
at $z \sim 3$, several times at $z \sim 4$, and more than a factor
of 10 at $z>4.6$ (the gray dashed-line traces the ratio of SDSS-DR3 to
the previous surveys).  The dip at $z\sim 4$
in the ratio of SDSS-DR3 to the previous surveys is largely due to 
the impressive high $z$ survey carried out by the Cambridge group.
We also stress that the increased sensitivity of the
SDSS-DR3 at $z>4$ should
be viewed conservatively; follow-up observations
should be performed to confirm the high redshift results.
In Figure~\ref{fig:gzsdss} we show the \gz\ curves for several cuts
on \snrlim\ and sample selection for the SDSS-DR3 quasars.   These cuts
are summarized in Table~\ref{tab:cuts} and will be referred
to throughout the text.

In contrast with the $g(z)$ curves for the \lya\ forest or
\ion{Mg}{2} systems \citep[e.g.][]{prochter05}, one notes no
significant features due to strong sky lines.  
This is because the sky lines can only lead to false positive detections
in our algorithm.  These are easily identified and ignored.

\begin{table}[ht]
\begin{center}
\caption{SDSS Cuts\label{tab:cuts}}
\begin{tabular}{lcccclc}
\tableline
\tableline
Label &SNR$_{lim}$& CLR$^a$ & n$_{DLA}$ \\
\tableline
SDSS-DR3\_4  & 4 & no & 525\\
SDSS-DR3\_4C & 4 & yes & 340\\
SDSS-DR3\_5  & 5 & no & 395\\
SDSS-DR3\_5C & 5 & yes & 183\\
SDSS-DR3\_8  & 8 & no & 155\\
SDSS-DR3\_8C & 8 & yes & 88\\
\tableline
\end{tabular}
\end{center}
\tablenotetext{a}{This entry refers to whether the quasar
sample adheres
to the color criteria described by \cite{richards02}.  In
practice, we perform the color-cut by only including quasars drawn
from plates 761 and greater.}
\end{table}

\section{\nhi\ Measurements}
\label{sec:dla}

\subsection{Damped \lya\ Candidates}

Damped \lya\ candidates were identified in the quasar spectra using
the same prescription introduced by PH04.  In short, the algorithm
compares the SNR of the pixels in a running window of width
$6(1+z)$\AA\ against $x_{SNR}$ specifically definied to be the 
median SNR of the 151 pixels starting 200 pixels redward of the \lya\ 
emission peak divided by 2.5, i.e.\ $x_SNR$ gauges the SNR of the
quasar continuum just redward of \lya\ emission.  Furthermore, 
$x_{SNR}$ is restricted to have a minimum value of 1 
and a maximum value of 2.
If the fraction of pixels with SNR$\geq x_{SNR}$ is $\leq 60\%$, a
damped \lya\ candidate is recorded at the center of the window.
This candidate list is supplemented by systems associated with
relatively strong metal-line transitions outside the \lya\ forest
(see Herbert-Fort et al.\ 2005, in preparation).
Finally, the list is further supplemented
by our visual inspection of each quasar spectrum
when characterizing its intrinsic
absorption.  The damped \lya\ candidates are
listed in Table~\ref{tab:qso}.

In PH04, we reported that Monte Carlo tests of our automated
algorithm on synthetic spectra implied an idealized
 damped \lya\ completeness
$\sim95\%$ for $\mnhi \approx 2\sci{20} \cm{-2}$ and $100\%$ for
$\log \mnhi  > 20.4$.  Given the supplemental candidates
from the metal-line search and our visual inspection, we believe
the completeness definitely exceeds $95\%$ for all absorbers with
$\mnhi  > 2\sci{20} \cm{-2}$.
We discuss a new set of completeness tests below.

Every damped \lya\ candidate was subjected to the following analysis.
First, the \lya\ profile was visually
inspected and obvious false-positive candidates were eliminated.  
This visual analysis included overplotting a \lya\ profile with 
$\mnhi  = \Nth$.  To minimize the labor of fitting \lya\ profiles,
we chose not to fit many systems where the overplotted 
profile clearly was a poor solution.  It was our experience that
nearly all of these candidates have best-fit values
$\mnhi  < 1.5 \sci{20} \cm{-2}$. 
Second, we searched for metal-line absorption at redshifts near 
the estimated redshift centroid of the \lya\ profile.  
The extensive wavelength coverage of
SDSS spectra is a tremendous advantage in the analysis of damped
\lya\ candidates.  In general, we focused on the strongest low-ion 
transitions observed in the damped \lya\ systems \citep[e.g.][]{pro03a}:
\ion{Si}{2}\,$\lambda 1260,1304,1526$, \ion{O}{1}\,$\lambda$1302, 
\ion{C}{2}\,$\lambda$1334,
\ion{Al}{2}\,$\lambda$1670, and \ion{Fe}{2}\,$\lambda$1608,2382,2600.
We characterized the metal-line absorption for each candidate
as: (a) no metals detected\footnote{We note that these damped \lya\
candidates may very well exhibit metal-line absorption at higher
spectral resolution and/or within the \lya\ forest.};
(b) weak or ambiguous metal absorption; and
(c) secure metal-line absorption.
Damped \lya\ candidates in the latter category generally exhibit
two or more metal-line features outside the \lya\ forest.  
For damped \lya\ systems with secure metal-line absorption, 
we constrain the subsequent \lya\ profiles to coincide with the
metal-line redshift.

\begin{figure}
\begin{center}
\includegraphics[height=3.6in,angle=90]{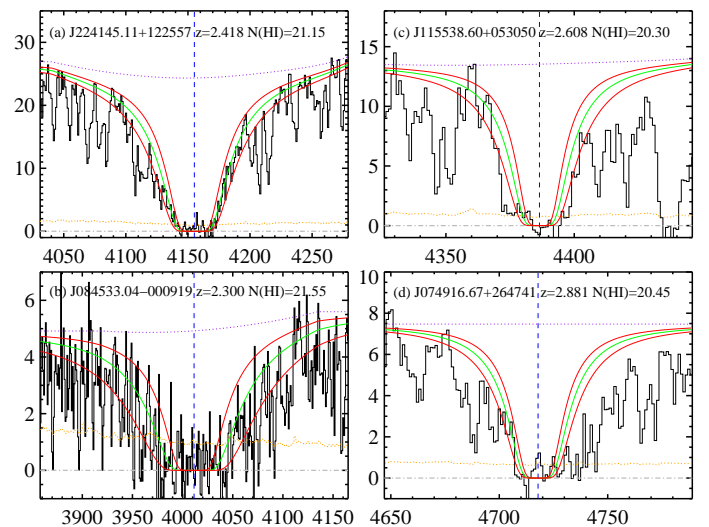}
\caption{Example profile fits to four damped \lya\ systems from 
our SDSS-DR3 survey.  The examples represent (a) a high SNR case
with a well constrained quasar continuum and minimal line blending
($\sigma(\mnhi) = 0.15$);
(b) a low SNR case ($\sigma(\mnhi) = 0.30$);
(c) an example with severe line blending ($\sigma(\mnhi) = 0.20$); and
(d) an example where determining the redshift from associated
metal-line absorption is very important ($\sigma(\mnhi) = 0.15$).
}
\label{fig:dlaex}
\end{center}
\end{figure}

\subsection{\lya\ Fits}

Those damped \lya\ candidates which were not rejected by visual inspection
were fitted with Voigt profiles in a semi-quantitative fashion.
As discussed by \cite{pro03a}, systematic error associated with 
continuum placement and line-blending generally dominates the
statistical error associated with Poisson noise in the quasar
flux.  We have fitted the \lya\ profile with an 
IDL tool {\it x\_fitdla} which allows the user to interactively
modify the Voigt profiles and continuum placements.  We then estimated a 
conservative uncertainty to each \nhi\ value based on the
constraints of the quasar continuum near the \lya\ profile,
the degree of line blending, and the Poisson noise of the data.
\cite{bolton04} have emphasized that the SDSS spectroscopic data
has non-Gaussian fluctuations, e.g. $3\sigma$
departures from the `true' value with higher than $0.27\%$ frequency.
Therefore, we have generally ignored single pixel outliers
when performing the \lya\ fits.

To be conservative, we adopt a minimum uncertainty of 0.15\,dex
in $\log \mnhi$
for the absorption systems with unambiguous metal-line detections
and 0.20\,dex otherwise.  
These uncertainties are larger than those reported in PH04.  As noted
above, the effects of line blending and continuum uncertainty dominate
the measurement uncertainty of the \nhi\ values.  Furthermore,
it is very unlikely
that the errors related to these systematic effects are Gaussian
distributed.  At present, it is very difficult to accurately assess
the magnitude and distribution of the errors in the \nhi\ values.
We discuss below an attempt based on the analysis of mock spectra,
yet even this approach has limited value.
Nevertheless, we contend that 95$\%$ of the true \nhi\ values will
lie within $2\sigma$ of the reported \nhi\ values, but we cannot
argue that the reported errors are Gaussian distributed.  
This remains a significant failing in the analysis of DLA surveys.

Figure~\ref{fig:dlaex} shows four example profile fits from
the SDSS sample which illustrate several of the key issues.  
In Figure~\ref{fig:dlaex}a we present a high signal-to-noise case with a 
well constrained quasar continuum and the redshift centroided by
metal-line detections.   The fit is constrained by data both
in the core and wings of the \lya\ profile and the adopted
0.15\,dex uncertainty is overly
conservative.   Figure~\ref{fig:dlaex}b presents a low 
signal-to-noise case.  In this case,
it is very difficult to constrain the fit
based on the core of the \lya\ profile.  Nevertheless, the continuum
level is reasonably well constrained and the wings of the profile
set the \nhi\ value.
Figure~\ref{fig:dlaex}c presents a case where line-blending is 
severe and the \nhi\ value is only constrained by the wings of the
profile.  In these cases the continuum
placement is critical.
Finally, Figure~\ref{fig:dlaex}d emphasizes the importance of metal-line
detection.  In this example, the \lya\ profile is centered on the 
redshift of the metal-line transitions.  Without this constraint,
we would have centered the \lya\ profile a few Angstroms redward
of the current position and derived a larger \nhi\ value.
We suspect that some damped \lya\ systems without
metal-line detections have $\mnhi $ values that are biased 
high ($\approx 0.1$ to 0.2\,dex).  Of course, these damped \lya\
systems tend to have lower $\mnhi $ values and this systematic
effect is primarily an issue only for candidates with 
$\mnhi  \approx \Nth$.
Finally, we should stress that the \nhi\ values of damped \lya\
candidates near the quasar \ion{O}{6} or \lya\ emission
features are difficult to constrain because of uncertain continuum
placement.


All of the fits from the DR3 sample and the \lya\ fits
from are presented at \\
http://www.ucolick.org/ $\sim$xavier/SDSSDLA/index.html.
In addition to postscript plots, we provide the files (IDL format)
which contain the \lya\ fits and quasar continua, and 
also a description of the software used to perform the fits.  Interested
readers can reanalyze any of our \lya\ fits.

\begin{table}[ht]\scriptsize
\begin{center}
\caption{{\sc SDSS-DR3 DLA STATISTICAL SAMPLE\label{tab:dlastat}}}
\begin{tabular}{lcccccc}
\tableline
\tableline
Quasar &$z_{abs}$ & DR & SNR & CLR & $f_{m}^b$& log \nhi \\
\tableline
J000238.4$-10$1149.8&3.2674&2&5&n&2&$21.20^{+0.15}_{-0.15}$\\
J000536.3$+13$5949.4&3.4896&2&5&n&2&$20.30^{+0.15}_{-0.15}$\\
J001115.2$+14$4601.8&3.6118&2&8&n&1&$20.60^{+0.25}_{-0.25}$\\
J001240.5$+13$5236.7&3.0217&2&8&n&2&$20.55^{+0.15}_{-0.15}$\\
J001328.2$+13$5827.9&3.2808&2&8&n&2&$21.55^{+0.15}_{-0.15}$\\
J001918.4$+15$0611.3&3.9710&2&5&n&2&$20.40^{+0.20}_{-0.20}$\\
J002614.6$+14$3105.2&3.3882&2&5&n&2&$20.70^{+0.15}_{-0.15}$\\
J003126.7$+15$0739.5&3.3583&2&8&n&2&$20.95^{+0.20}_{-0.20}$\\
J003501.8$-09$1817.6&2.3376&1&8&n&1&$20.55^{+0.15}_{-0.15}$\\
J003749.1$+15$5208.3&3.4790&2&4&n&1&$20.35^{+0.20}_{-0.20}$\\
\tableline
\end{tabular}
\end{center}
\tablenotetext{a}{0=No metals; 1=Weak metals; 2=Metals detected}
\tablecomments{[The complete version of this table is in the electronic\\
edition of the Journal.  The printed edition contains only a sample.]}
\end{table}
 
\begin{table}[ht]\scriptsize
\begin{center}
\caption{SDSS-DR3 DLA NON-STATISTICAL SAMPLE\label{tab:dlanon}}
\begin{tabular}{lcccc}
\tableline
\tableline
Name &$z_{abs}$ & DR & $f_{mtl}^a$ & log \nhi \\
\tableline
J001115.2$+14$4601.8&3.4523&2&2&$21.65^{+0.20}_{-0.20}$\\
J001134.5$+15$5137.4&4.3175&2&1&$20.50^{+0.20}_{-0.20}$\\
J001134.5$+15$5137.4&4.3592&2&2&$21.10^{+0.20}_{-0.20}$\\
J004950.9$-09$3035.6&3.2858&2&2&$20.70^{+0.20}_{-0.20}$\\
J014214.7$+00$2324.3&3.3482&1&2&$20.40^{+0.15}_{-0.15}$\\
J020651.4$-09$4141.3&2.4702&2&2&$20.30^{+0.20}_{-0.20}$\\
J021232.1$-10$0422.1&2.7140&1&0&$21.00^{+0.15}_{-0.15}$\\
J023148.8$-07$3906.3&2.2982&3&2&$21.75^{+0.20}_{-0.20}$\\
J033344.4$-06$0625.1&3.9349&3&2&$21.60^{+0.20}_{-0.20}$\\
J073149.5$+28$5448.7&2.6859&2&1&$20.55^{+0.15}_{-0.15}$\\
\tableline
\end{tabular}
\end{center}
\tablenotetext{a}{0=No metals; 1=Weak metals; 2=Metals detected}
\tablecomments{[The complete version of this table is in the electronic\\
edition of the Journal.  The printed edition contains only a sample.]}
\end{table}

\begin{table}[ht]\scriptsize
\begin{center}
\caption{SDSS-DR3 SUPER-LLS SAMPLE\label{tab:superlls}}
\begin{tabular}{lcccccc}
\tableline
\tableline
Quasar &$z_{abs}$ & DR & SNR & CLR & $f_{m}^b$& log \nhi \\
\tableline
J001328.2$+13$5828.0&2.6123&3&5&n&1&$20.10^{+0.15}_{-0.15}$\\
J002614.6$+14$3105.2&3.9039&2&5&n&2&$20.10^{+0.15}_{-0.15}$\\
J004732.7$+00$2111.3&2.4687&2&5&n&2&$20.00^{+0.15}_{-0.15}$\\
J013317.7$+14$4300.3&2.4754&2&4&n&2&$20.00^{+0.20}_{-0.20}$\\
J013317.7$+14$4300.3&2.9766&2&8&n&2&$20.15^{+0.15}_{-0.15}$\\
J014609.3$-09$2918.2&3.6804&2&8&n&1&$20.25^{+0.20}_{-0.20}$\\
J021143.3$-08$4723.8&2.2684&2&4&y&2&$20.10^{+0.15}_{-0.15}$\\
J021232.1$-10$0422.1&2.2738&1&8&n&2&$20.25^{+0.15}_{-0.15}$\\
J025039.1$-06$5405.1&4.3894&3&4&n&2&$20.00^{+0.15}_{-0.15}$\\
J031036.8$+00$5521.7&3.1142&2&5&n&2&$20.20^{+0.15}_{-0.15}$\\
\tableline
\end{tabular}
\end{center}
\tablenotetext{a}{0=No metals; 1=Weak metals; 2=Metals detected}
\tablecomments{[The complete version of this table is in the electronic\\
edition of the Journal.  The printed edition contains only a sample.]}
\end{table}

Tables~\ref{tab:dlastat} to \ref{tab:superlls} present the \nhi\ 
values for the damped \lya\ systems comprising the statistical sample,
the damped \lya\ systems discovered which are not in any of our 
statistical samples, and all of the
damped \lya\ candidates with central \nhi\ values less than 
$\Nth$ (termed super-LLS).  Regarding Table~\ref{tab:dlastat}, 
we list the damped \lya\ absorption redshift, indicate its membership
within the various statistical cuts, list the data release of the actual
spectrum analyzed\footnote{In general this corresponds to the first
data release when the quasar was observed by SDSS although errors
in our bookkeeping 
may have delayed its analysis.}, present the metal-line
characteristics, and list the \nhi\ value and error.
The non-statistical damped \lya\ sample is comprised of damped \lya\
systems with $z_{abs} \approx z_{qso}$, $z_{abs} < z_i$, and/or
systems toward quasars with strong intrinsic absorption.  We 
present this Table for completeness.  We emphasize that while
the absorbers listed in Table~\ref{tab:superlls} have 
a central value below $\Nth$, the true \nhi\ value 
of many of these absorbers will exceed the damped \lya\ threshold.

\begin{figure}[ht]
\begin{center}
\includegraphics[height=3.6in,angle=90]{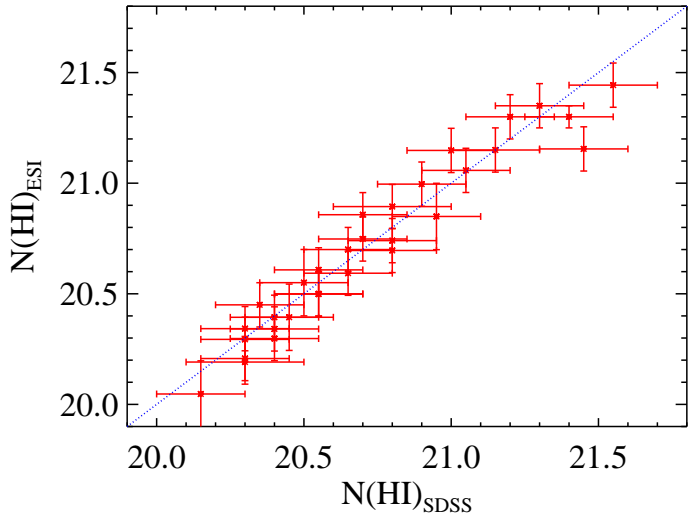}
\caption{Comparison of the \nhi\ values for a sub-set of the SDSS-DR3
sample as measured from the SDSS quasar spectra and, independently,
data acquired with the Echellette Spectrometer and Imager.
}
\label{fig:allcmp}
\end{center}
\end{figure}

\subsection{Completeness Tests and \nhi\ Accuracy}
\label{sec:acc}

There are several tests one can perform to assess the reliability
and completeness of the \nhi\ analysis.  
A valuable test of the \nhi\ accuracy is to compare the SDSS
measurements against values derived from observations at
higher spectral resolution and/or SNR.  Over the past several years,
we have observed a subset of the SDSS damped \lya\ sample
for other scientific studies \citep[e.g. chemical evolution and
\ion{C}{2}$^*$ analysis;][]{pro03b,wolfe04} with the 
Echellette Spectrometer and Imager \citep[ESI;][]{sheinis02}.  
Figure~\ref{fig:allcmp} compares the values
of the SDSS fits with those derived from the ESI spectra.
The agreement is excellent.  
It is important to note, however, that 
the fits were performed primarily by one of us (JXP) and that the systematic
error related to continuum placement is likely correlated.
Similarly, the effects of line-blending are qualitiatively similar
for the two data sets.
Indeed, the reduced $\chi^2_\nu = 0.4$
indicates the estimated uncertainties significantly
exceed the statistical uncertainty.   There is a notable case
from PH04 (at $z=2.77$ toward J084407.29+515311) where the SDSS value
is 0.3\,dex higher than the ESI value due to improper continuum 
placement in the SDSS analysis.  We believe this one case was
the result of our initial inexperience with the SDSS spectra
and that such errors are very rare in the current analysis.
Nevertheless, it does
underscore the fact that the error in the \nhi\ values
are predominantly systematic and in some cases large.

\begin{figure}[ht]
\begin{center}
\includegraphics[height=3.6in,angle=90]{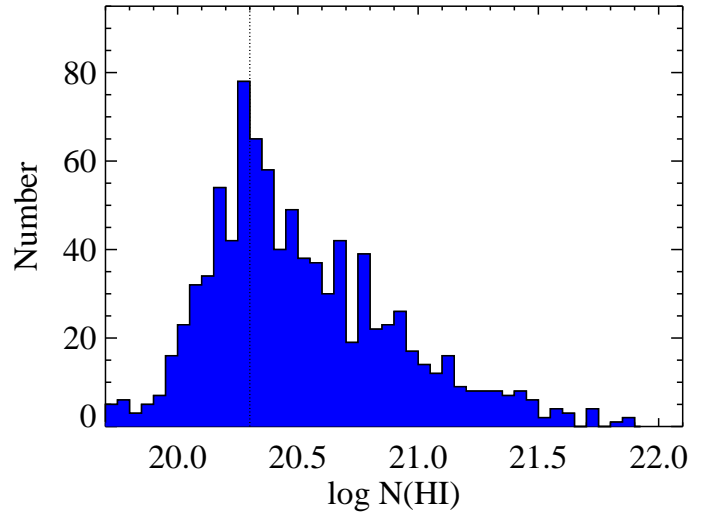}
\caption{Histogram of the \nhi\ values for every damped \lya\
candidate fit in the SDSS-DR3 sample.  Note that the distribution
peaks at $\mnhi < \Nth$ suggesting a high level completeness at
the damped \lya\ threshold (dotted line).  We currently estimate
the completeness level to be $\approx 99\%$.
}
\label{fig:hihist}
\end{center}
\end{figure}

One qualitative assessment of sample completeness is an inspection of 
the \nhi\ distribution.
Figure~\ref{fig:hihist}
presents a histogram of \nhi\ values for all of the fitted damped \lya\
candidates.  
The distribution shows a steady increase to lower \nhi\ value which
extends just below $\Nth$.
We note that
the relatively steep decline in the distribution at $\mnhi < 10^{20.2} \cm{-2}$
is partly due to incompleteness in our search algorithm.  It is also
due to the fact that we eliminated many damped \lya\ candidates after 
visual inspection. 

Since the publication of PH04, we have performed a damped \lya\ survey
on a new set of mock spectra kindly provided by P. McDonald.  These mock
spectra were carefully constructed to match the SNR and redshifts of the
full SDSS-DR3 quasar sample.  A synthetic \lya\ forest was added to the
quasar continuum according to the prescriptions detailed in 
\cite{mcdonald05a}.  Finally, a random set of 
damped \lya\ systems (with a column density frequency distribution similar
to the results below) were added to the spectra and also a set
of Lyman limit systems with $\mnhi = 1\sci{17}$ to $2\sci{20} \cm{-2}$
following the prescriptions of \cite{mcdonald05b}.
We analyzed these mock spectra using the same algorithms as the 
scientific search and recovered 99$\%$ of the damped \lya\
systems.  The few that we did not `discover'
have $\mnhi < 10^{20.4} \cm{-2}$ and 
just missed satisfying the damped \lya\ candidate criteria.
Furthermore, we recovered 90$\%$ of the absorbers with 
$\mnhi \approx 1\sci{20}\cm{-2}$.  Therefore, we are confident that the
sample is nearly complete although it is likely we are missing $\approx 5\%$
of the absorbers with $\mnhi \approx 2\sci{20} \cm{-2}$.

To further investigate uncertainty in our \nhi\ measurements, we 
fitted Voigt profiles to 50 of the damped \lya\ systems from the mock
spectra.   With only one exception, our fitted values are within 
0.3~dex of the true value.  In this one 
exception we fitted one damped \lya\ profile to a case that was actually
two damped systems separated by $\approx 500 \mkms$.  In practice, this 
situation can be accounted for if both damped \lya\ systems exhibit
metal-line absorption.  In any case, we expect these systems to be 
rare \citep{lopez03},
but a non-zero systematic error to the damped \lya\ analysis.
In future papers, we will expand this type of mock analysis to probe
other aspects of fitting errors and completeness limits for the SDSS samples.

\section{RESULTS}
\label{sec:HI}

In this section, we report the main results of the damped \lya\
survey.  These include the \ion{H}{1} frequency distribution
function and its zeroth and first moments.  It is amusing to note
that many of the techniques are directly analogous to
the derivation and analysis of galaxy luminosity functions.
Indeed, we rely on many of the standard techniques and face 
similar challenges and systematic uncertainties.

\begin{figure*}
\begin{center}
\includegraphics[width=5.3in]{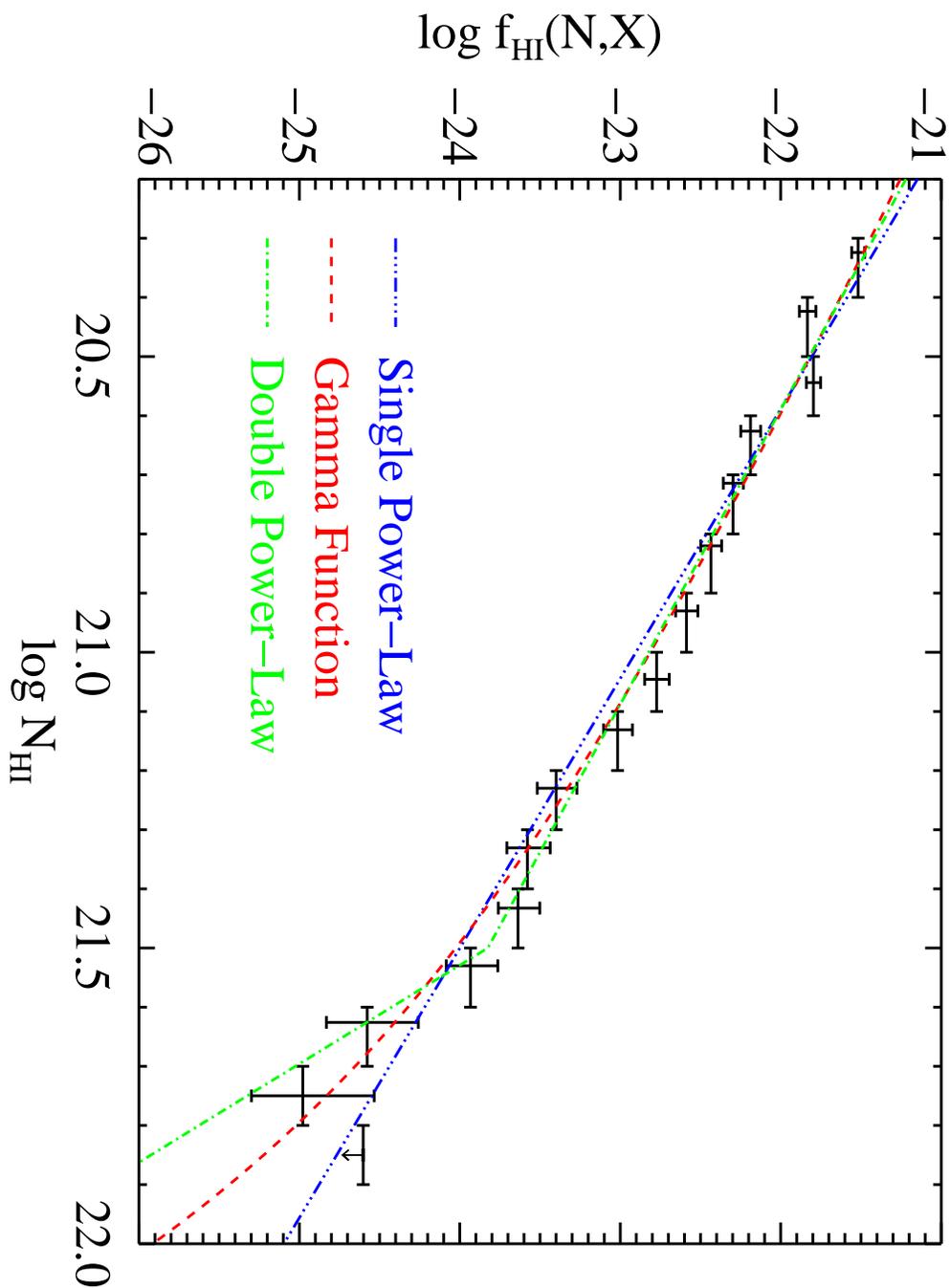}
\caption{The \ion{H}{1} 
frequency distribution \fnhi\ for all of the damped \lya\
systems identified in the SDSS-DR3\_4 sample (mean redshift $z=3.06$).   
Overplotted on the
discrete evaluation of \fnhi\ are the fits of a single power-law,
a $\Gamma$-function, and a double power-law.  Only the latter two
are acceptable fits to the observations.
}
\label{fig:fnall}
\end{center}
\end{figure*}

\clearpage

\subsection{\fnhi: The \ion{H}{1} Column Density Frequency Distribution}
\label{sec:fn}

Following previous work \citep[e.g.][]{lzwt91}, we define the 
number of damped \lya\ systems
in the intervals $(N,N+dN)$ and $(X,X+dX)$:

\begin{equation}
f_{\rm HI}(N,X) dN dX \cmma
\end{equation}

\noindent where \fnhi\ is the frequency distribution
and the `absorption distance'

\begin{equation}
dX \equiv \frac{H_0}{H(z)} (1+z)^2 dz \perd
\end{equation}

\noindent  With this definition, an unevolving population of objects
will have constant \fnhi\ in time provided one adopts the
correct cosmology.

In Figure~\ref{fig:fnall}, we present \fnhi\ for the 
full SDSS-DR3\_4 survey.  This sample spans the redshift interval
$z=[2.2, 5.5]$ with an integrated absorption pathlength
$\Delta X = 7417.5$ and a \nhi-weighted mean redshift of 3.2.
The solid points with error bars describe \fnhi\
for \nhi\ bins of $\Delta N = 0.1$\,dex: 

\begin{equation}
f_{\rm HI}(N,X) = \frac{m_{DLA} (N,N+\Delta N)}{\Delta X} \cmma
\end{equation}

\noindent  with $m_{DLA}$ the number of damped \lya\ systems
within $(N, N+\Delta N)$ in the $\Delta X$ interval.
The error bars reflect Poisson uncertainty (68.3$\%$ c.l.)
according to the value of $m_{DLA}$ and the upper limits
correspond to 95$\%$\,c.l. 
The derivation of \fnhi\ in this discrete manner is analogous
to the derivation of galaxy luminosity functions using the $V_{max}$
method.

Lacking a physical model, 
we have considered three functional forms to describe \fnhi:
(i) a single power-law

\begin{equation}
f_{\rm HI}(N,X) = k_1 N^{\alpha_1} \cmma
\label{eqn:sngl}
\end{equation}

\noindent (ii) a $\Gamma$-function \citep[e.g.][]{fall93}

\begin{equation}
f_{\rm HI}(N,X) = k_2 \ltp \frac{N}{N_\gamma} \rtp^{\alpha_2} 
\exp \ltp \frac{- N }{N_\gamma} \rtp \cmma
\end{equation}

\noindent and (iii) a double power-law

\begin{equation}
f_{\rm HI}(N,X) = k_3 \ltp \frac{N}{N_d} \rtp^\beta 
  \; {\rm where} \; \beta =  
\begin{cases}
\alpha_3:  N < N_d ; \quad \\ 
\alpha_4:  N \geq N_d \\
\end{cases}
\label{eqn:dbl}
\end{equation}

\noindent  In each case, we perform a maximum likelihood analysis 
to constrain the functional parameters
and set the normalization constants ($k_1, k_2, k_3$) by imposing
the integral constraint

\begin{equation}
\intl_{N_{th}}^\infty f(N,X) dN = \frac{m_{DLA}}{\Delta X}  \perd
\end{equation}

To estimate the parameter uncertainties and construct a correlation
matrix, we performed a `jack-knife' analysis 
\citep[e.g.][]{lupton93,blanton03}.
Specifically, we derived the best fit parameters for 21~subsamples ignoring 222
random quasars in each case.  The parameter uncertainty is then given
by

\begin{equation}
\sigma^2 = \frac{N-1}{N} \smm_i \ltp x_i - \bar x \rtp^2
\end{equation}

\noindent with $N=21$.  
In some cases whese values are close to the one-parameter confidence limits
derived from the maximum likelihood function although there are notable
exceptions.
We calculate the correlation matrix in standard
fashion \citep[see][]{blanton03},  

\begin{equation}
r_{ij} = \frac{\avgq{\Delta x_i \Delta x_j}}{\ltp \avgq{\Delta x_i^2}
\avgq{\Delta x_j^2} \rtp^\ohf}
\end{equation}

with

\begin{equation}
\avgq{\Delta x_i \Delta x_j} = \frac{N-1}{N} \smm_i 
(x_i - \bar x_i) (x_j - \bar x_j)  \perd
\end{equation}

\begin{table*}
\begin{center}
\caption{FITS TO $f_{\rm HI}(N,X)$\label{tab:fnfits}}
\begin{tabular}{cccccccc}
\tableline
\tableline
Form &Parameters
& $z \epsilon [2.2,5.5)$\tablenotemark{a}
& $z \epsilon [1.7,2.2)$
& $z \epsilon [2.2,2.5)$
& $z \epsilon [2.5,3.0)$
& $z \epsilon [3.0,3.5)$
& $z \epsilon [3.5,5.5)$ \\
\tableline
Single & $\log k_1$ &$23.16^{+0.02}_{-0.02}$&$21.81^{+0.08}_{-0.08}$&$27.15^{+0.04}_{-0.04}$&$22.12^{+0.03}_{-0.03}$&$20.84^{+0.03}_{-0.03}$&$23.33^{+0.04}_{-0.04}$\\
& $\alpha_1$ &$-2.19^{+0.05}_{-0.05}$&$-2.13^{+0.18}_{-0.23}$&$-2.39^{+0.13}_{-0.15}$&$-2.14^{+0.08}_{-0.08}$&$-2.08^{+0.08}_{-0.09}$&$-2.20^{+0.10}_{-0.11}$\\
Gamma & $\log k_2$ &$-23.52^{+0.02}_{-0.02}$&$-23.03^{+0.08}_{-0.08}$&$-23.40^{+0.04}_{-0.04}$&$-23.70^{+0.03}_{-0.03}$&$-23.03^{+0.03}_{-0.03}$&$-23.99^{+0.04}_{-0.04}$\\
& $\log N_\gamma$ &$21.48^{+0.09}_{-0.06}$&$21.31^{+0.35}_{-0.15}$&$21.31^{+0.29}_{-0.12}$&$21.56^{+0.16}_{-0.10}$&$21.36^{+0.12}_{-0.08}$&$21.69^{+0.14}_{-0.14}$\\
& $\alpha_2$ &$-1.80^{+0.06}_{-0.06}$&$-1.56^{+0.13}_{-0.27}$&$-1.94^{+0.16}_{-0.16}$&$-1.78^{+0.09}_{-0.10}$&$-1.52^{+0.09}_{-0.11}$&$-1.93^{+0.12}_{-0.12}$\\
Double & $\log k_3$ &$-23.83^{+0.02}_{-0.02}$&$-24.09^{+0.08}_{-0.08}$&$-24.16^{+0.04}_{-0.04}$&$-23.69^{+0.03}_{-0.03}$&$-23.51^{+0.03}_{-0.03}$&$-23.76^{+0.04}_{-0.04}$\\
& $\log N_d$ &$21.50^{+0.05}_{-0.04}$&$21.65^{+0.15}_{-0.08}$&$21.50^{+0.18}_{-0.07}$&$21.45^{+0.12}_{-0.05}$&$21.45^{+0.09}_{-0.06}$&$21.50^{+0.20}_{-0.07}$\\
& $\alpha_3$ &$-2.00^{+0.06}_{-0.06}$&$-1.96^{+0.23}_{-0.26}$&$-2.25^{+0.16}_{-0.17}$&$-1.94^{+0.10}_{-0.10}$&$-1.81^{+0.10}_{-0.11}$&$-2.03^{+0.12}_{-0.13}$\\
& $\alpha_4$ &$ -6.00^{+4.06}_{-3.93}$&$-10.00^{+8.27}_{-7.78}$&$-10.00^{+7.91}_{-7.59}$&$ -4.26^{+2.41}_{-2.22}$&$ -5.72^{+4.02}_{-3.81}$&$ -4.78^{+2.87}_{-2.63}$\\
\tableline
\end{tabular}
\end{center}
\tablenotetext{a}{Restricted to the SDSS-DR3\_4 sample.  The remaining columns include the full sample.}
\tablecomments{The errors reported are one-parameter errors which 
do not account for correlations among the parameters. 
See Table~\ref{tab:corrmat} for the correlation matrix.}
\end{table*}
 
\begin{table*}
\begin{center}
\caption{CORRELATION MATRIX FOR $f_{\rm HI}(N,X)$\label{tab:corrmat}}
\begin{tabular}{ccccccccccc}
\tableline
\tableline
Param &$\sigma$ &$\delta \log k_1$ &$\delta \alpha_1$ &$\delta \log k_2$ 
&$\delta \log N_\gamma$ &$\delta \alpha_2$ &$\delta \log k_3$ &$\delta \log N_d$
&$\delta \alpha_3$ &$\delta \alpha_4$ \\
\tableline
$\delta \log k_1$ &  1.04&$  1.00$&$ -1.00$&$ -0.48$&$  0.28$&$ -0.80$&$ -0.33$&$ -0.01$&$ -0.93$&$ -0.06$\\
$\delta \alpha_1$ &  0.05&$ -1.00$&$  1.00$&$  0.48$&$ -0.29$&$  0.80$&$  0.32$&$  0.02$&$  0.93$&$  0.05$\\
$\delta \log k_2$ &  0.29&$ -0.48$&$  0.48$&$  1.00$&$ -0.98$&$  0.91$&$  0.47$&$ -0.25$&$  0.66$&$ -0.28$\\
$\delta \log N_\gamma$ &  0.10&$  0.28$&$ -0.29$&$ -0.98$&$  1.00$&$ -0.80$&$ -0.45$&$  0.29$&$ -0.50$&$  0.29$\\
$\delta \alpha_2$ &  0.11&$ -0.80$&$  0.80$&$  0.91$&$ -0.80$&$  1.00$&$  0.48$&$ -0.17$&$  0.89$&$ -0.15$\\
$\delta \log k_3$ &  0.18&$ -0.33$&$  0.32$&$  0.47$&$ -0.45$&$  0.48$&$  1.00$&$ -0.93$&$  0.61$&$  0.53$\\
$\delta \log N_d$ &  0.08&$ -0.01$&$  0.02$&$ -0.25$&$  0.29$&$ -0.17$&$ -0.93$&$  1.00$&$ -0.29$&$ -0.64$\\
$\delta \alpha_3$ &  0.07&$ -0.93$&$  0.93$&$  0.66$&$ -0.50$&$  0.89$&$  0.61$&$ -0.29$&$  1.00$&$  0.09$\\
$\delta \alpha_4$ &  2.13&$ -0.06$&$  0.05$&$ -0.28$&$  0.29$&$ -0.15$&$  0.53$&$ -0.64$&$  0.09$&$  1.00$\\
\tableline
\end{tabular}
\end{center}
\tablecomments{Restricted to the SDSS-DR3\_4 sample.}
\end{table*}

\noindent As is typically the
case for fits to galaxy luminosity functions, we find significant correlation
between the parameters.  It is important to keep this in mind when
comparing the fits to various redshift intervals.
Table~\ref{tab:fnfits} presents the best-fit
values for the parameters for the full SDSS-DR3 sample and also 
the fits to the damped \lya\ systems in several redshift intervals.
In this table, the error bars refer to `one-parameter uncertainties' that
correspond to the 68$\%$ c.l. of the maximum likelihood function when
keeping the other parameters fixed at their best-fit values.
The exceptions are the normalization
values where we have only reported the Poissonian error 
based on the best fit.
Table~\ref{tab:corrmat} provides the parameter uncertainties 
and the correlation matrix for the full SDSS-DR3\_4 sample.
When the absolute value of the off-diagonal terms is much less than 1,
then there is little correlation between the parameters.
Unfortunately, there are too few damped \lya\ systems in the redshift 
bins to perform a meaningful jack-knife analysis for these subsets.
We suspect, however, that the parameters are correlated in a similar
way as to the results presented in Table~\ref{tab:corrmat}.

The best-fit solutions are overplotted on the binned evaluation of
\fnhi\ in Figure~\ref{fig:fnall}.  
First, consider the single power-law solution (dotted line) with a best
fit slope of $\alpha_1 = -2.19 \pm 0.05$.
This functional form is a poor description of the data.
A one-sided Kolmogorov-Smirnov (KS) test indicates there
is a less than $0.1\%$ probability 
that the cumulative distributions of the observations and
the power-law are drawn from the same parent population.
In short, the power-law is too steep at low \nhi\ and too shallow 
at large \nhi.   Although previous surveys suggested
a single power-law was a poor description \citep[e.g.][]{wolfe95}, 
their sample size was too
small to rule out this solution.  
Note that this result contrasts the damped \lya\ systems with the \lya\ forest
(absorbers with $\mnhi < 10^{15} \cm{-2}$)
where a single power-law with exponent $\alpha_1 \approx -1.5$ is a good
description of the observations \citep[e.g.][]{kirkman97}.

Although a single power-law is a poor description of the observations,
the fit does highlight an important new result: the \fnhi\ distribution
is steeper than a $\mnhi^{-2}$ power-law at large column density.
We will further develop this point in $\S$~\ref{sec:omg}.
The other two curves on Figure~\ref{fig:fnall} show the 
$\Gamma$-function (dashed line) and double power-law fit (dash-dot line).
Both of these functional forms are a fair fit to the observations;
a one-sided KS test gives values $>10\%$ for the full 1$\sigma$ range
of the parameters.
Furthermore, there is good agreement between
the `break' column
densities ($N_\gamma$ and $N_d$) and the power-law 
indices at low column densities are consistent.  Both 
solutions indicate that \fnhi\ drops very steeply ($\alpha \ll -2$) at
$\mnhi \approx 10^{21.5} \cm{-2}$
and that the distribution has a `faint-end' slope of $\alpha \approx -2$.

\begin{figure}[ht]
\begin{center}
\includegraphics[height=3.6in,angle=90]{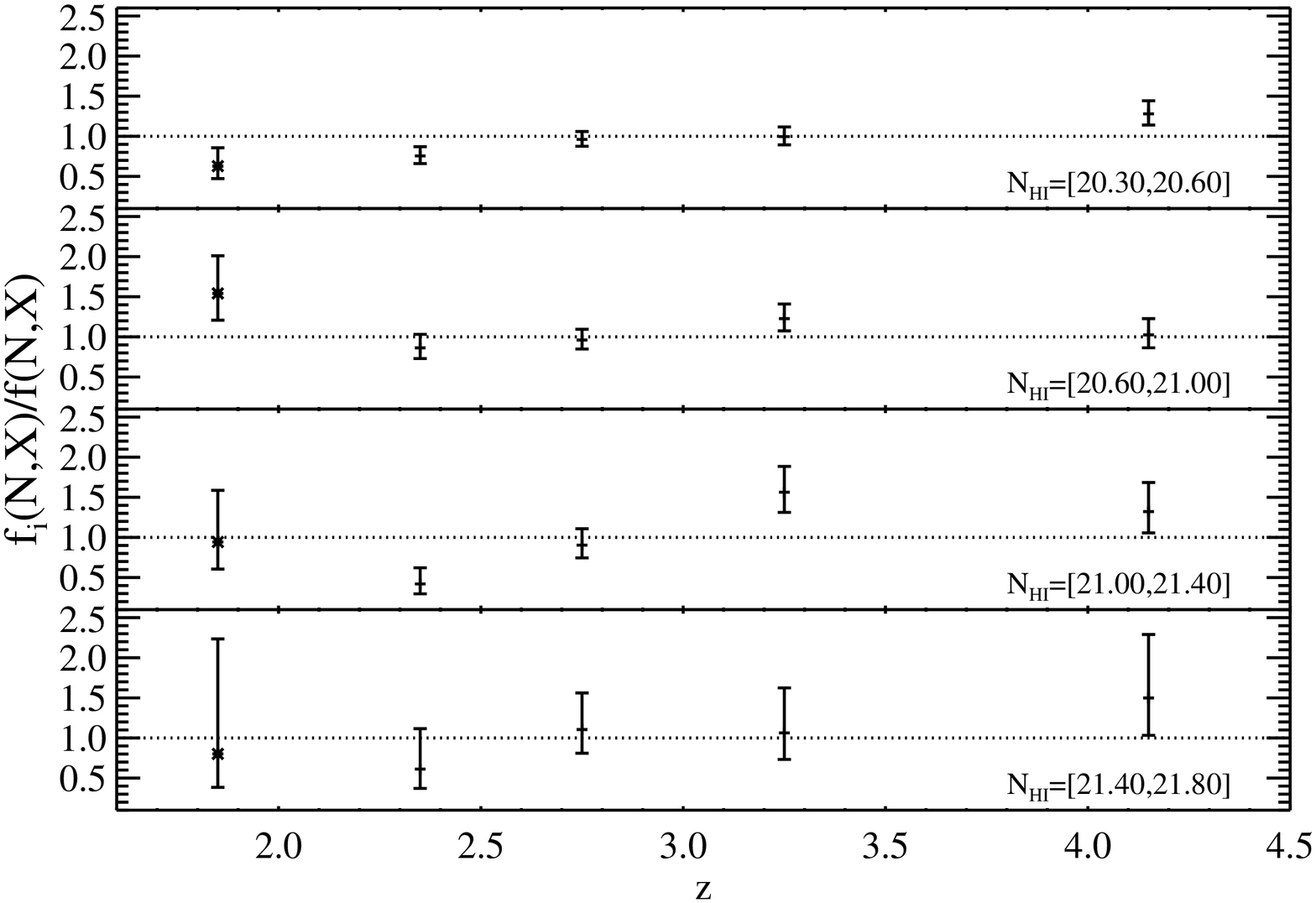}
\caption{Plot of the \nhi\ frequency distribution of damped \lya\
systems in 5 redshift intervals against the full distribution.
The four \nhi\ intervals are:
(1) $20.3 \leq \log \mnhi < 20.6$;
(2) $20.6 \leq \log \mnhi < 21.0$;
(3) $21.0 \leq \log \mnhi < 21.4$;
and
(4) $21.4 \leq \log \mnhi < 21.8$.
One observes significant evolution in \fnhi\ at low \nhi\ value but
that the shape of \fnhi\ is nearly invariant with redshift.
Finally, note that the data 
points for $z<2.2$ (marked with a cross) do not include
measurements from the SDSS survey.
}
\label{fig:fncomp}
\end{center}
\end{figure}

In the following sub-sections, we will consider evolution in the
zeroth and first moments of \fnhi.  
Figure~\ref{fig:fncomp} qualitatively describes
the redshift evolution of the full 
\fnhi\ distribution.  The figure plots \fnhi\ as a function of
redshift against the \fnhi\ distribution
of the full SDSS-DR3 sample in four \nhi\ intervals: 
(1) $20.3 \leq \log \mnhi < 20.6$;
(2) $20.6 \leq \log \mnhi < 21.0$;
(3) $21.0 \leq \log \mnhi < 21.4$;
and
(4) $21.4 \leq \log \mnhi < 21.8$.
At low \nhi, 
\fnhi\ increases monotonically with redshift.
The peak-to-peak range is modest, however, only evolving by 
a factor of $\approx 2$.
One identifies similar evolution in the high \nhi\ bins with the 
exception of the highest redshift interval.  

Perhaps the most 
important result is that 
the shape of \fnhi\ is independent of redshift.
We have compared the \fnhi\ distributions from each redshift bin
using a two-sided
Komolgorov-Smirnov (KS) test.  This test compares the shape of the 
distributions but is insensitive to the normalization.  Our analysis
found KS probabilities greater than $10\%$ for every pair of 
redshift intervals.  
Contrary to previous claims, therefore, there is no evidence for a
significant evolution in \fnhi\ for the DLAs with 
redshift (e.g.\ a steeping beyond
$z=3$).  With the current sample, 
there may only be significant evolution in the 
normalization of \fnhi.

\begin{figure}
\begin{center}
\includegraphics[height=3.6in,angle=90]{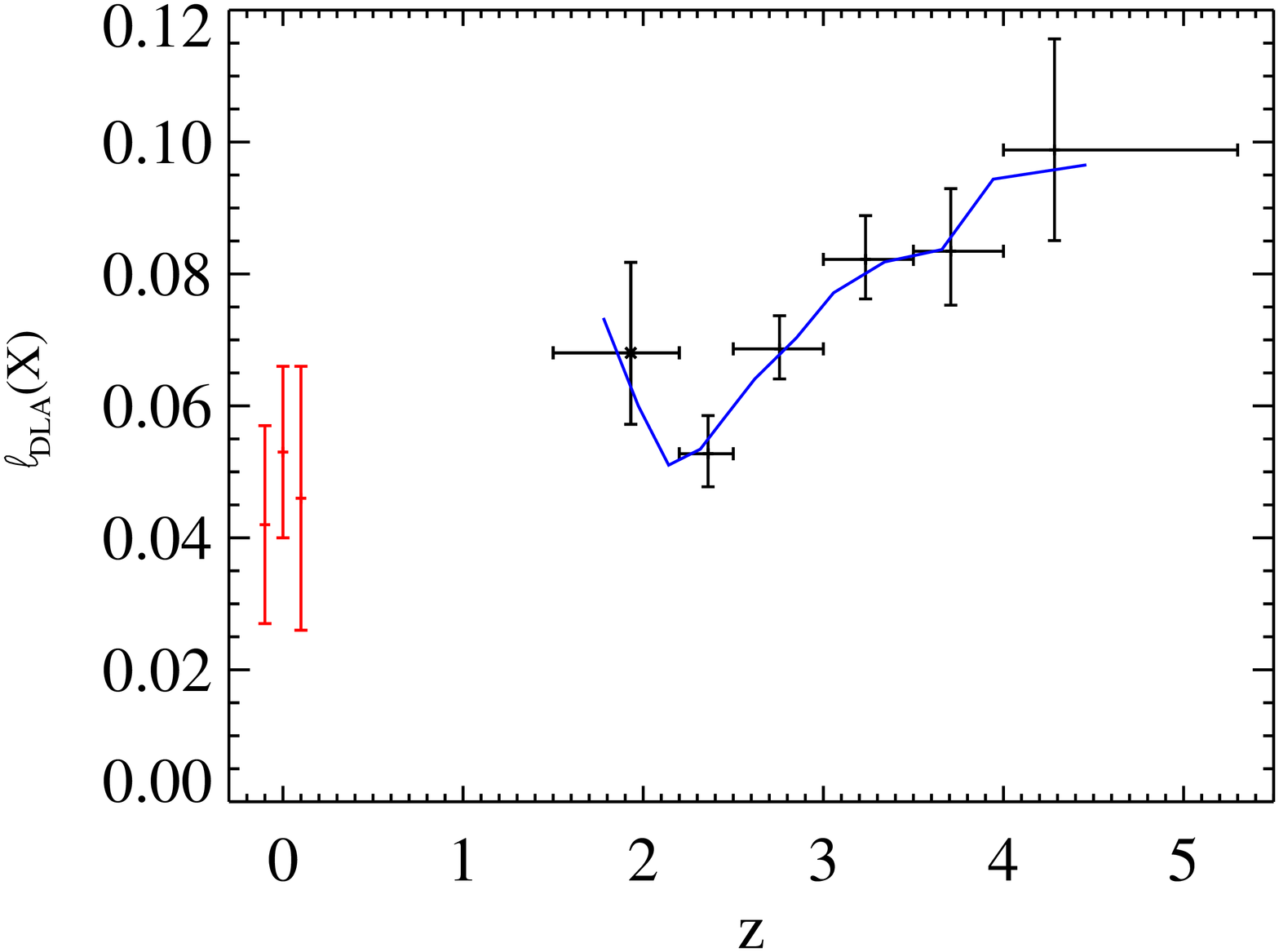}
\caption{Plot of the line density of damped \lya\ systems $\ldla(X)$
versus redshift.  The points at $z=0$ are from three 
analyses of 21cm observations by \cite{zwaan01,rosenbergs03,ryan03,ryan05}.
The curve overplotted on the data 
traces the evolution of $\ldla(X)$ in a series of 0.5\,Gyr 
intervals.  Contrary to previous studies (which focused on the line
density in redshift space $\ldla(z)$), we find statistically significant
evolution in the line density per unit absorption distance.
}
\label{fig:taux}
\end{center}
\end{figure}

\subsection{$\ldla (X)dX$: The Incidence of Damped \lya\ Systems}

The zeroth moment of \fnhi\ gives the number of damped \lya\ systems
encountered per unit absorption pathlength $dX$, i.e.\ the line-density
of damped \lya\ systems\footnote{Previous work has represented
this quantity with various terminology that is often confused with the
column density or number density (e.g.\ $dN/dX, dn/dz$).}:

\begin{equation}
\ldla(X)dX = \intl_{N_{th}}^\infty \mfnhi dN dX \perd
\end{equation}
\noindent The line density is related to the comoving number density of 
damped \lya\ systems $n_{DLA}(X)$ and the cross-section
$A(X)$ as follows:

\begin{equation}
\ldla(X) = (c/H_0) n_{DLA}(X) A(X) \perd
\label{eqn:cover}
\end{equation}
\noindent  In this manner, $\ldla$ is related to the covering fraction
of damped \lya\ systems on the sky.

Figure~\ref{fig:taux} shows $\ldla(X)dX$ for the damped \lya\ systems
at $z>1.6$ and for the local universe 
\citep{zwaan01,rosenbergs03,ryan03,ryan05}.
The values for the damped \lya\ systems were calculated in the 
discrete limit, e.g.\

\begin{equation}
\ldla(X) = \frac{m_{DLA}}{\Delta X}
\end{equation}
\noindent and the error bars reflect Poisson uncertainty in 
$m_{DLA}$ ($68.3\%$ c.l.).   The solid line traces the
value of $\ldla(X)$ derived in a series of 0.5\,Gyr time intervals
to reveal any differences in binning.

It is evident from the figure that the line density of damped \lya\
systems increases from $z=2$ to 4;
the increase between the $z=[2.2,2.5]$ interval and the 
$z=[3.5,5,5]$ bin is a factor of $1.7 \pm 0.2$.  The largest change
in $\ldla(X)$ occurs during a $\Delta z = 1$ interval from
$z \approx 2.3$ to 3.3 corresponding to $\Delta t = 800$\,Myr.
We discuss possible explanations for this evolution in $\S$~\ref{sec:discuss}.
Another interesting result is that $\ldla(z\approx 2.3$) is
less than $2\sigma$ larger than $\ldla(z=0)$, i.e.\ 
the data suggest little evolution in the line density of damped systems
over the past $\approx 10$Gyr.
Within the context of hierarchical cosmology,
this is a surprising result.  It implies that when smaller galaxies merge
and accrete other systems, the product of the comoving density
and the total gas cross-section satisfying the
damped \lya\ criterion is conserved.  
Perhaps this is
related to the fact that the damped \lya\ threshold is near the
surface density threshold for star formation.

\subsection{\omg: The Cosmological Neutral Gas Mass Density}
\label{sec:omg}

\subsubsection{Definition of \omg}

The first moment of the \fnhi\ distribution gives a cosmologically
meaningful quantity, the gas mass density of \ion{H}{1} atoms:

\begin{equation}
\Omega^{\rm HI}_g(X) dX \equiv \frac{\mu m_H H_0}{c \rho_c} 
  \intl_{N_{min}}^{N_{max}} \mnhi \mfnhi \, dX
\label{eqn:omg}
\end{equation}

\noindent where $\mu$ is the mean molecular mass of the gas (taken to be 1.3),
$H_0$ is Hubble's constant, and $\rho_c$ is the critical mass density.
Traditional treatments of \ohi\ and the damped \lya\ systems have
integrated Equation~\ref{eqn:omg} from $N_{min}=\Nth$ to $N_{max} = \infty$
yielding \odla.
As emphasized by \cite{peroux03}, \odla\ may be significantly less than
\ohi\ if absorbers below the damped \lya\ threshold contribute to the
\ion{H}{1} mass density.  
In this case, \ohi\ would include a large contribution from gas
which is predominantly ionized because the \ion{H}{1} atoms in the
Lyman limit systems are a mere tracer of the gas.  
It is difficult, however, to assign any physical significance to \ohi\
aside from a mere census of the \ion{H}{1} atoms in the universe.
In contrast, \odla\ offers a
good estimate of the mass density of gas which is predominantly
neutral (see below).

In the following, we will consider both quantities with
primary emphasis on \odla\ because we contend
it best defines the gas reservoir available for star formation at high
redshift.
In practice, one generally evaluates \odla\ in
the discrete limit

\begin{equation}
\Omega_g^{\rm DLA} = \frac{\mu m_H H_0}{c \rho_c} 
\frac{\Sigma \mnhi}{\Delta X} \;\; ,
\label{eqn:omgdisc}
\end{equation}
\noindent where the sum is performed over the \nhi\ measurements
of the damped \lya\ systems in a given redshift interval with
total pathlength $\Delta X$.

In the following, we will consider several definitions for
\omg\ based on the values of $N_{min}$ and $N_{max}$.
Table~\ref{tab:omg} summarizes the various definitions.

\begin{table}[ht]\footnotesize
\begin{center}
\caption{DEFINITIONS OF \omg \label{tab:omg}}
\begin{tabular}{cccl}
\tableline
\tableline
Def. &$N_{min}$ 
& $N_{max}$ & Description \\
& ($\cm{-2}$) & ($\cm{-2}$) \\
\tableline
\ohi & 0 & $\infty$ & Density of gas associated with
\ion{H}{1} atoms \\
$\momg^{Neut}$ & ?? & $\infty$ & Density of predominantly
neutral gas \\
\odla & $2 \sci{20}$ & $\infty$ & Density of
gas associated with the DLA \\
\olls & $1.6 \sci{17}$ & $2 \sci{20}$ 
& Density of the gas associated with LLS \\
$\momg^{21cm}$ & --- & --- & Density of \ion{H}{1} gas 
in collapsed objects \\
\tableline
\end{tabular}
\end{center}
\end{table}

\subsubsection{Convergence of \omg}

Consider first the upper limit, $N_{max}$.   To verify \omg\
converges, it
is necessary to integrate \fnhi\ until 
\begin{equation}
\frac{d \log \mfnhi}{d \log \mnhi} \ll -2  \cmma
\label{eqn:conv}
\end{equation}

\begin{figure}
\begin{center}
\includegraphics[height=3.6in,angle=90]{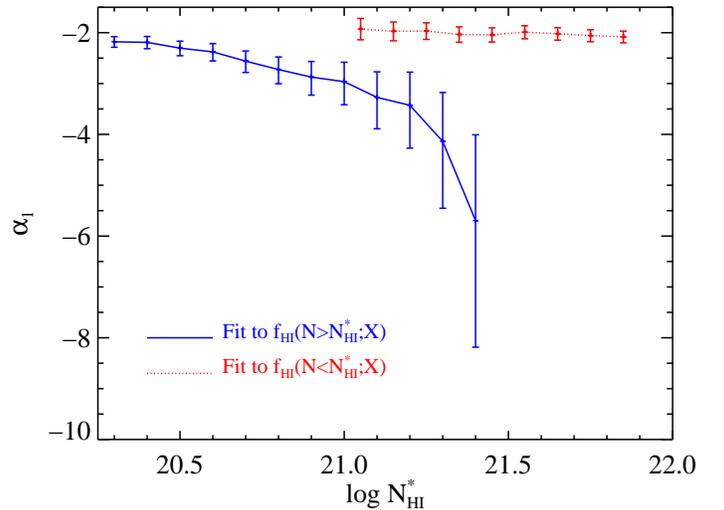}
\caption{The solid line plots the maximum likelihood value 
of the exponent for a single power-law fit to the \fnhi\ distribution
of the SDSS-DR3\_4 sample as a function of the minimum \nhi\ value
of the distribution.  The value is less than $-2$ for all values of
$\mnhi^*$ and the curve indicates the power-law steepens as 
one restricts the analysis to the high \nhi\ end of the distribution
function.   The dotted curve, meanwhile, plots the exponent value
as a function of the maximum \nhi\ value of the \fnhi\ distribution.
There is little evolution in this value although the slope is 
more shallow at the low \nhi\ region of the distribution.
}
\label{fig:fnconv}
\end{center}
\end{figure}

\noindent i.e.\ until one establishes that
\fnhi\ is significantly steeper than a $\mnhi^{-2}$ power-law.
As noted in $\S$~\ref{sec:fn} and Figure~\ref{fig:fnall}, the 
$\Gamma$-function and double power-law fits
to the \fnhi\ distribution indicate \fnhi\ steepens
at \nhi~$\approx 10^{21.5} \cm{-2}$.  Furthermore, even
a single power-law fit to the data almost
satisfies  Equation~\ref{eqn:conv}.  
This point is emphasized in Figure~\ref{fig:fnconv}
which presents the exponent of the
best fitting single power-law to the SDSS-DR3 \fnhi\ distribution
function as a function of the lower \nhi\ bound to the distribution
function, e.g., the point at $\log \mnhi^* = 21$ shows the exponent
for the single power-law fit to $f_{\rm HI}(N_{\rm HI} > 10^{21} \cm{-2};X)$.
Not surprisingly, the curve starts at $\mnhi = \Nth$ with 
$\alpha_1 = -2.2$ and decreases with increasing $\mnhi^*$.   
In addition, it is important
that the value of $\alpha_1$ decreases below $-3$ in
Figure~\ref{fig:fnconv} because a power-law 
with $\alpha_1 = -2.2$ converges very slowly with $N_{max}$.

\begin{figure}[ht]
\begin{center}
\includegraphics[height=3.6in,angle=90]{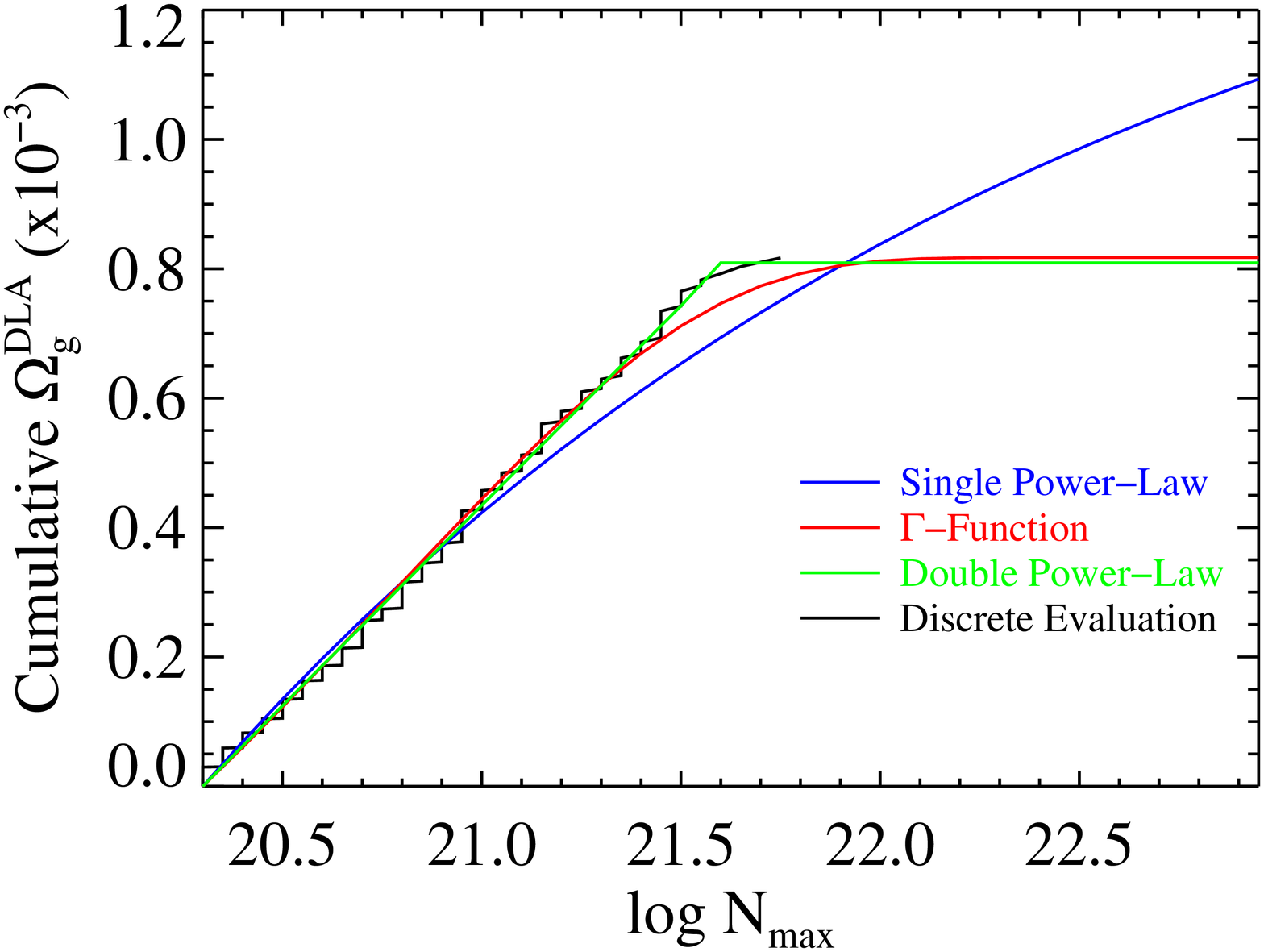}
\caption{Cumulative evaluation of \odla\ for the SDSS-DR3\_4
sample as a function of the maximum \nhi\ value.  
The black curves trace the evaluation of \odla\ in the discrete
limit (Equation~\ref{eqn:omgdisc}) while the smooth curves
show the evaluation by integrating the functional fits to the
\fnhi\ distribution.  Both the double power-law and $\Gamma$-function
converge to the discrete value by $\mnhi = 10^{22} \cm{-2}$.  An
extrapolation of the single power-law, however, has not converged
even at $\mnhi = 10^{23} \cm{-2}$.
}
\label{fig:omgcon}
\end{center}
\end{figure}

Figure~\ref{fig:omgcon} shows the cumulative \odla\ value against
$\log N_{max}$.  Both the double power-law and
$\Gamma$-function converge
to the value indicated by the discrete
evaluation of \odla\ (Equation~\ref{eqn:omgdisc}).
This adds additional confidence
that these functional forms are good representations of \fnhi.
The single power-law, however, has not converged at the highest
\nhi\ value observed for a damped \lya\ system; it implies
significantly larger \odla\ values than currently supported by the data. 
Nevertheless, even this functional form converges formally as
$N_{max} \to \infty$: $\alpha_1 < -2$ at $3\sigma$ significance.
{\it The results presented here are the first definitive
demonstration that \odla\ converges and most likely 
by $\mnhi \approx 10^{22} \cm{-2}$.}
We emphasize that smaller redshift intervals
(i.e.\ subsets of the SDSS-DR3 sample) have insufficient
sample size to convincingly argue for convergence.  Therefore,
one may cautiously view the \odla\ values presented as a function
of redshift.

Now consider the lower bound to the integrand, $N_{min}$.
One evaluates \ohi\ with $N_{min} = 0$,
i.e.\ including the contribution from all classes of quasar
absorption line systems.   
We have argued that this definition for \omg\ lacks
special physical significance, yet let us first consider its
value.  In particular, we will investigate the contribution of 
\odla\ to \ohi.
\cite{tytler87} first demonstrated that the \ion{H}{1} frequency
distribution of the \lya\ forest (absorbers with $\mnhi < 10^{17} \cm{-2}$)
is well described by a single
power-law with exponent $\alpha_{\rm Ly\alpha} \approx -1.5$.
More recent studies give
similar results over the \nhi\ interval $10^{12} - 10^{15} \cm{-2}$
\citep[e.g.][]{hu95,kirkman97}. 
This functional form is sufficiently shallow that the contribution
of the \lya\ forest to \ohi\ is negligible in comparison to
absorbers with $\mnhi > 10^{15} \cm{-2}$.

\subsubsection{Impact of the Lyman Limit Systems}

Of greater interest to this discussion is the contribution of the
Lyman limit systems\footnote{Lyman limit systems are 
defined to be all absorbers with
optical depth exceeding unity at 1\,Ryd.  In this discussion,
we generally restrict the definition to be 
absorbers with $17.2 \leq \log (\mnhi/\cm{-2}) < 20.3$.
The one exception is the line density, as 
defined in Equation~\ref{eqn:llsline}.} 
to \ohi.
Indeed, \cite{peroux03} claimed that Lyman limit systems with
$\mnhi > 10^{19} \cm{-2}$ contribute $\approx 50\%$ of \ohi\ at
$z>3.5$.   In PH04, we argued that \cite{peroux03} overstated the
contribution of these `super-LLS' to \ohi\ and, in particular,
underestimated the uncertainty of their contribution.
Let us return to this issue with the increased sample of SDSS-DR3.

\begin{figure}[ht]
\begin{center}
\includegraphics[height=3.6in,angle=90]{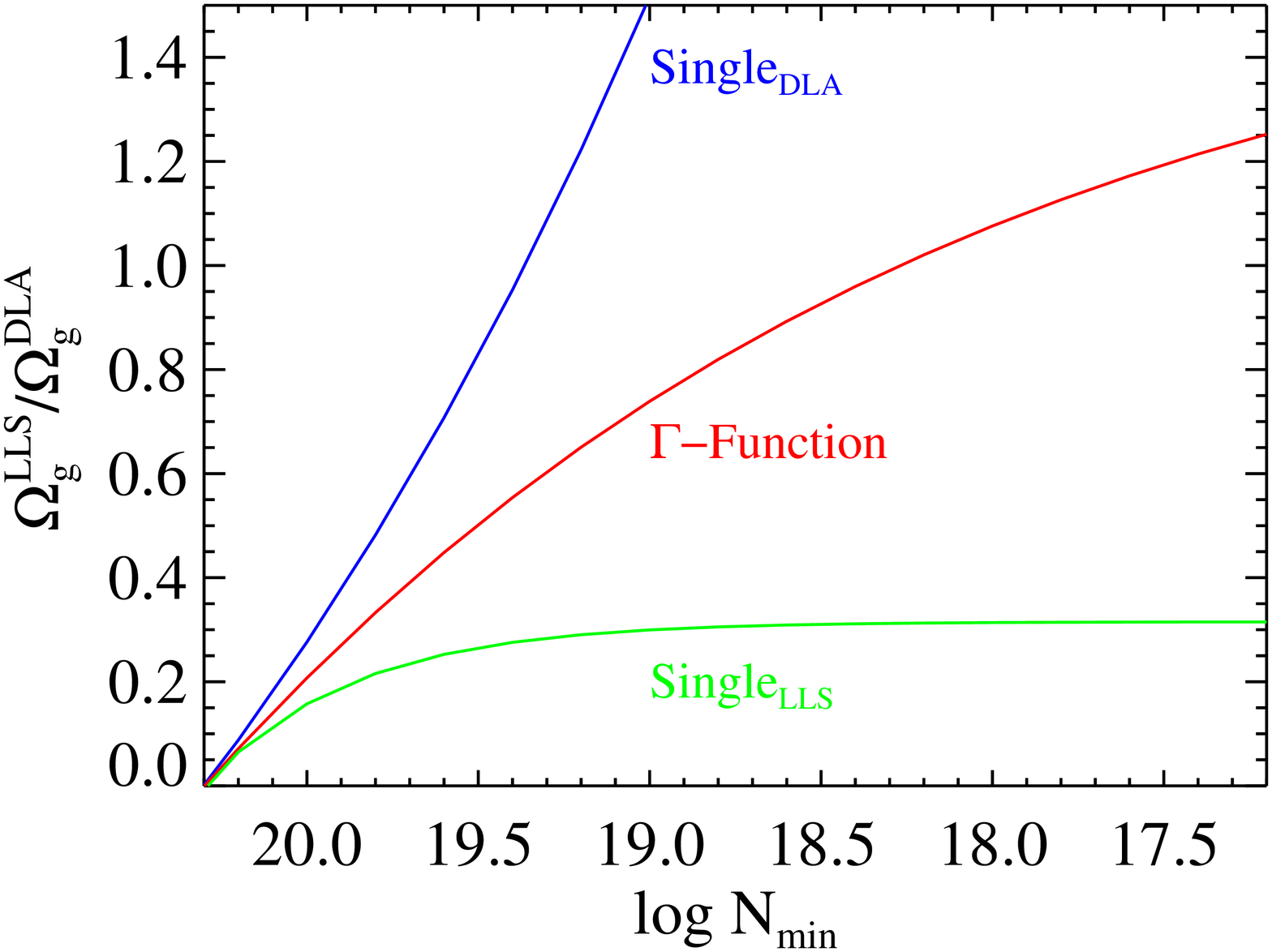}
\caption{Contribution to \ohi\ of the Lyman limit
systems relative to the damped \lya\ systems as a function of the
minimum \nhi\ value of the Lyman limit systems.
The top two curves refer to extrapolations of the single power-law
and $\Gamma$-function fits to \fnhi\ for the damped \lya\ systems.
The single-power law is divergent and even the $\Gamma$-function
has not converged at $10^{17.2} \cm{-2}$.  These extrapolations,
however, significantly overpredict the line density of Lyman limit systems
and should be considered upper limits to 
$\Omega_g^{\rm LLS}/\Omega_g^{\rm DLA}$.  The lower curve is 
the result for a single power-law fit to \fnhi\ in the \nhi\ 
interval for Lyman limit systems.  Because the first derivative
of this single power law is not continuous at the damped \lya\
system threshold or the transition to the \lya\ forest, it should
be considered a lower limit to $\Omega_g^{\rm LLS}/\Omega_g^{\rm DLA}$.
}
\label{fig:cumomg}
\end{center}
\end{figure}

The first point to stress is that 
$d \log \mfnhi / d \log \mnhi \approx -2$ at 
$\mnhi \approx 10^{20.3} \cm{-2}$.  This is sufficiently steep that
absorbers with $\mnhi \approx 10^{20} \cm{-2}$ must contribute
at least a few percent of \ohi.  Indeed, if we extrapolate 
our fits to \fnhi\ for the damped \lya\ systems from 
$\mnhi = 2 \sci{20} \cm{-2}$ to $10^{17} \cm{-2}$,
we find the Lyman limit systems would give a larger
neutral gas mass density than
the value of the damped \lya\ systems.  
Figure~\ref{fig:cumomg}
presents the contribution of the Lyman limit systems to \omg\
relative to the damped \lya\ systems $\Omega_g^{\rm LLS}/\Omega_g^{\rm DLA}$.
The upper curves show the results for extrapolations of the
the single power-law and the $\Gamma$-function fits to \fnhi\ for
the damped \lya\ systems.
For the single power-law, the gas mass density diverges as 
$N_{min} \to 0$ because $\alpha_1 < -2$.  Even with 
the $\Gamma$-function (with 
exponent $\alpha_2 = -1.8$) the Lyman limit systems 
dominate the contribution to \ohi.  

There is, however, an 
additional observational constraint which limits \olls/\odla.
Although the functional form
of \fnhi\ is essentially unconstrained for the Lyman limit systems,
their line density,

\begin{equation}
\ell_{LLS}(X) \equiv \ell(N>10^{17.2} \cm{-2};X) \cmma
\label{eqn:llsline}
\end{equation}

\noindent has been measured at $z>2$
\citep{peroux01}.  The extrapolations of the single power-law and
$\Gamma$-function fits to the damped \lya\ frequency distribution
overpredict the observed value of the incidence
of Lyman limit systems by more than an order of magnitude.
Therefore, the estimate of \olls\ for these curves
should be considered unrealistic upper limits.

We can set a lower bound to \olls\ by assuming \fnhi\ is a 
single power-law over the $10^{17.2} \cm{-2} < \mnhi < 10^{20.3} \cm{-2}$
interval:

\begin{equation}
\flls(N,X) = k_{\rm LLS} \ltp \frac{N}{10^{20.3} \cm{-2}} 
\rtp^{\alpha_{\rm LLS}} \perd
\end{equation}
\noindent  The two parameters of this function are determined by
demanding

\begin{equation}
\flls(10^{20.3} \cm{-2}; X) = 
f_{\rm HI}^{\rm DLA}(10^{20.3} \cm{-2}; X) 
\end{equation}

\noindent and

\begin{equation}
\intl_{10^{17.2}}^{10^{20.3}} \flls dN \, dX = 
\ell_{\rm LLS} dX - \ell_{\rm DLA} dX  \perd
\end{equation}

\noindent To estimate $\ell_{\rm LLS}(X)$, we adopt the functional
form provided by \cite{peroux03} as a function of redshift
$\ell_{\rm LLS}(z) = 0.07 (1+z)^{2.45}$, and adopt the mean
value for the entire SDSS-DR3 survey, i.e. 

\begin{equation}
\ell_{\rm LLS}(X) = \frac{\int \ell_{\rm LLS}(z) g(z) dz}{\Delta X} \perd
\end{equation}

The uncertainty in $\ell_{\rm LLS} (z)$ implies an approximately $20\%$
uncertainty in $\ell_{\rm LLS} (X)$.  To estimate 
$\fdla(N=10^{20.3} \cm{-2})$ and determine $k_{\rm LLS}$, 
we adopt the central value of our
$\Gamma$-function fit.  The relative contribution of 
\olls/\odla, of course, is relatively insensitive 
to this parameter.  For the full
SDSS-DR3\_4 sample, we find $\log k_{\rm LLS} = -21.43$
and $\alpha_{\rm LLS} = -1.01^{+0.09}_{-0.02}$.
It is not coincidental that this slope matches the best-fit
values of the `faint-end' slope for the $\Gamma$-function fits
to \fnhi\ performed by \cite{storrie00} and \cite{peroux03}.
Their maximum likelihood analyses included the integral constraint
imposed by the Lyman limit systems and the greater than
three orders of magnitude lever arm in \nhi\ forces this result.

Because the LLS 
power-law is more shallow than the `faint-end' slope of the
fits to $\fdla$,  the estimate of \olls\ based on
$\flls$ must be considered a lower limit.  
The \olls/\odla\ curve for the single-power law fit to
$\flls$ is also shown in Figure~\ref{fig:cumomg}.
We find \olls~$> 0.3 \,$\odla\
and that the majority of the contribution comes from absorbers
with $\mnhi > 10^{19} \cm{-2}$.
As noted above, \cite{peroux03} first cautioned that the super-LLS
could contribute significantly to \ohi, especially at very high redshift.
In PH04, we argued that \cite{peroux03} had overstated the effect at $z>3.5$
and underestimated the uncertainty.
Our new results indicate the super-LLS contribute 
from 20 to $50\%$ to \ohi\ for the complete SDSS-DR3 survey.
Also, recall that the shape of \fnhi\ is statistically invariant with
redshift (Figure~\ref{fig:fncomp}).  Therefore, it is likely that
the Lyman limit systems offer a non-negligible
contribution to \ohi\ at all redshifts.
To determine the value of \olls\ as a function of redshift,
one compares the
relative evolution of \llls\ and $\ldla$. 
Current observations suggest \llls\ has a stronger redshift
dependence than $\ldla$.  
In fact, we do find \olls\ (as determined from the single
power-law fit to $\flls$) to increase with redshift (Table~\ref{tab:summ}).
There is a comparable increase in \odla, however, and the
ratio \olls/\odla\ is roughly constant in time.

\begin{figure}[ht]
\begin{center}
\includegraphics[height=3.6in,angle=90]{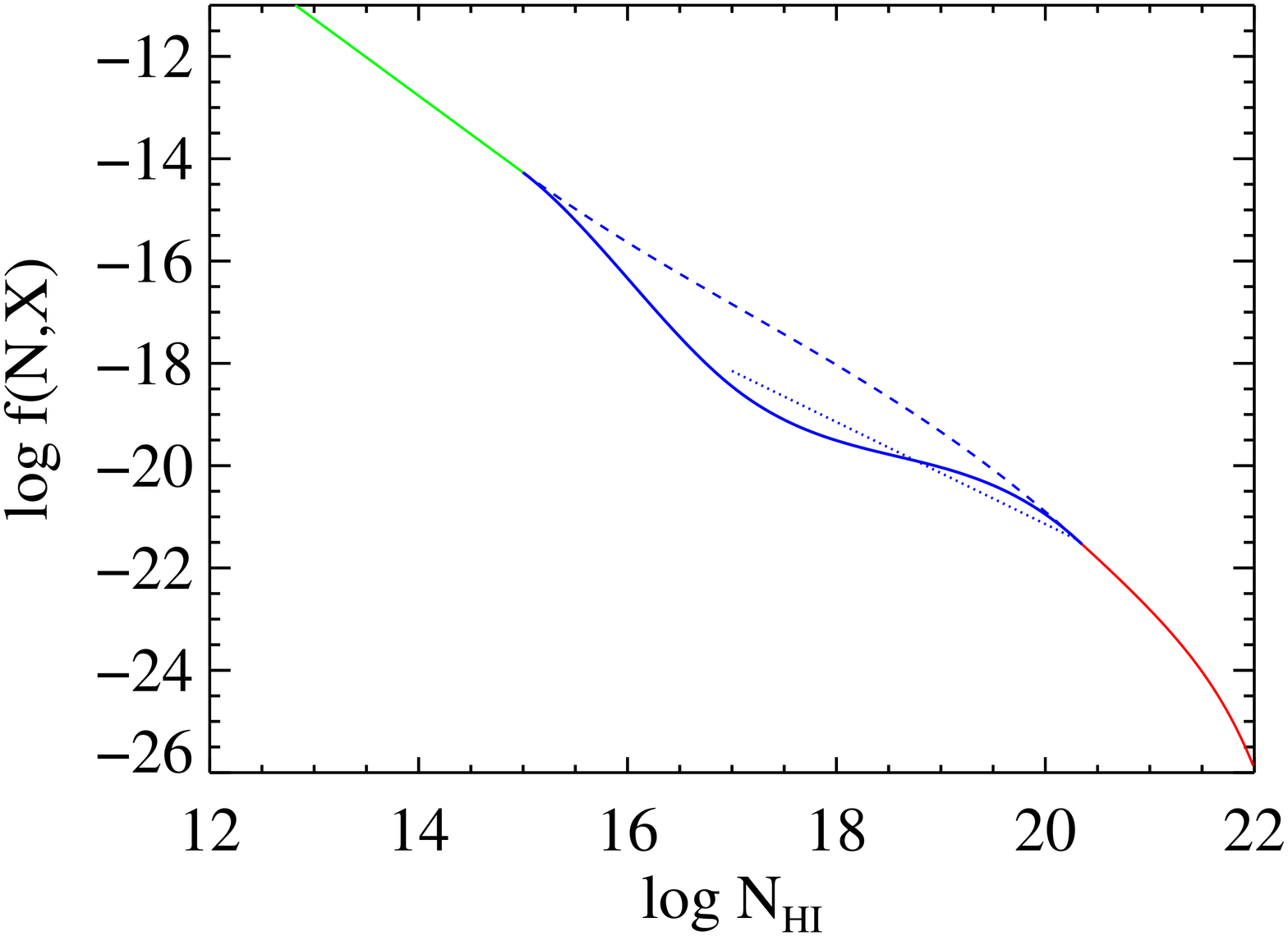}
\caption{Figure depicting the \fnhi\ distribution of the quasar
absorption line systems at $z=2.7$.  
The distribution for the 
damped \lya\ systems corresponds to the $\Gamma$-function 
fit to the SDSS-DR3\_4 sample.  The distribution for the \lya\
forest ($\mnhi < 10^{15} \cm{-2}$) is taken from 
\cite{kirkman97} scaled to the $\Lambda$CDM cosmology.  
We show three curves for the Lyman limit
system regime which should only be considered as a sketch:
(1) the dotted curve is the single power-law fit to $\flls$.
It is constrained to match the incidence of Lyman limit systems
at $z \sim 2.7$;
(2) the dashed curve is a spline function constrained
to match the \lya\ forest and damped \lya\ systems.
It clearly overpredicts the incidence of Lyman limit systems;
and (3) the solid curve is a spline function fit to the \lya\
forest and damped \lya\ systems and also constrained by a single
point at $\mnhi = 10^{17} \cm{-2}$ defined such that the integral
constraint for the Lyman limit systems is satisfied.
}
\label{fig:fncartoon}
\end{center}
\end{figure}

We reemphasize that the value of \olls\ derived from the single-power
law fit to $\flls$ is an underestimate of the true value.
In Figure~\ref{fig:fncartoon}, we show a spline solution which matches the
\fnhi\ functional forms of the \lya\ forest at $\mnhi = 10^{15} \cm{-2}$
and the damped \lya\ systems at $\mnhi = \Nth$ and also gives the
correct line density of Lyman limit systems.
We can give an estimate of the LLS contribution based on this spline
function: \olls/\odla~$= 0.57$.  
Therefore, the likely contribution of the Lyman limit systems to 
\ohi\ is $\approx 50\%$\,\odla\ and they comprise $\approx 1/3$ of \ohi.
In $\S$~\ref{sec:discuss},
we consider an alternate definition of the gas mass density:
the mass density of gas that is predominantly neutral, \oneut.  
Under this definition, one
may disregard the majority of the Lyman limit systems
on the grounds that they are highly ionized 
\citep{pro96,pro99,pz99,dessauges03},
but we suspect that at least a subset with $\mnhi \approx 10^{20} \cm{-2}$
is significantly neutral and could feed star formation.
This neutral subset, however, may contribute only a few 
percent relative to \oneut\ and we contend that the damped \lya\
systems contain $> 80\%$ of \oneut\ at high redshift.
Until a robust determination of \fnhi\ and the ionization
state of the gas in the Lyman limit systems
is made, the only conservative practice is to 
restrict the discussion of \ohi\ to include only those regions of the
universe with surface density $\mnhi \geq \Nth$
(i.e.\ \odla).
For the near future, this is the only quantity that will be
precisely measured.

As an aside, consider the implications of the 
$\alpha_{\rm LLS}$ value.  
First, this value is significantly more shallow than
the power-law dependence of the \lya\ forest.  Second, 
it is shallower than the `faint-end' slope of the damped \lya\
\fnhi\ distribution.   If we demand that \fnhi\ and its
first derivative are continuous across the \lya\ forest/LLS and
LLS/DLA boundaries, then our results indicate that $\flls$ has an
inflection with $d \log f_{\rm HI}^{\rm LLS}/d \log N > -1$ 
(Figure~\ref{fig:fncartoon}).
It is not surprising that $\flls$ would have an unusual
functional form given that it spans the regime from highly ionized
gas to primarily neutral gas.  
In fact, the inflection indicated by our analysis resembles
the functional forms suggested by 
\cite{zheng02} and \cite{maller03}.  
In passing, we advertise that a principal goal of
the MIKE/HIRES Lyman Limit Survey is to study $\flls$
and the ionization state of the Lyman limit systems
(Prochaska, Burles, Prochter, O'Meara, Bernstein, \& Schectman, in prep.).

\subsubsection{\odla\ Evolution}
\label{sec:omgevo}

\begin{figure}[ht]
\begin{center}
\includegraphics[height=3.6in,angle=90]{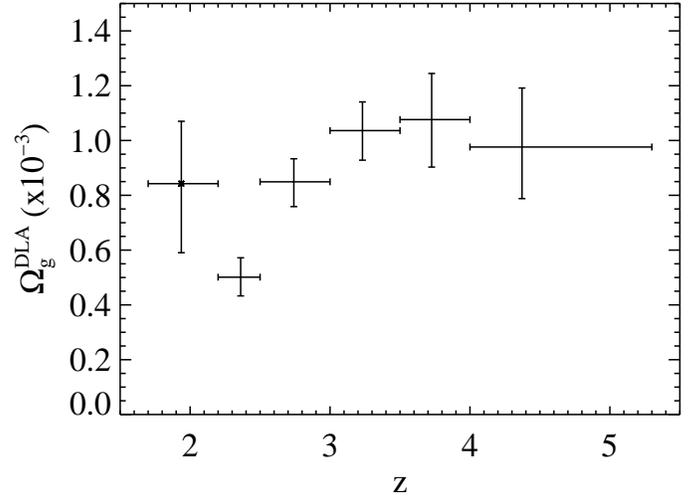}
\caption{Neutral gas mass density of
the damped \lya\ systems alone as a function of redshift.
There is an increase of approximately a factor of 
two in \odla\ from $z=2$ to 3
with the majority of rise occurring in only 500\,Myr.
Note the data point at $z<2.2$ (marked with a cross) does not include
measurements from the SDSS survey.
}
\label{fig:omega}
\end{center}
\end{figure}

Let us consider\footnote{The reader should refer to $\S$~\ref{sec:qsomag}
for a discussion of an important systematic effect related to \odla. 
For the remainder of this paper, we present results which ignore the
effect.  Our present concern is that we may be overestimating \odla.}
the redshift evolution of \odla\ 
by evaluating Equation~\ref{eqn:omgdisc} in a series of
redshift intervals.  Figure~\ref{fig:omega} presents the results.
Following PH04, we estimate the uncertainty in \odla\ 
through a modified bootstrap technique.  Specifically, 
we perform 5000 trials where we randomly draw $p$ damped \lya\
systems from the observed sample
where $p$ is a normally distributed random number with
mean equal to the number of damped systems in the interval $(m_{DLA})$
and with variance also equal to $m_{DLA}$.  We calculate \odla\
for each trial and calculate the upper and lower values 
corresponding to $68.3\%$ of the distribution.

The results presented in Figure~\ref{fig:omega} present the
first statistically significant evidence that \odla\ evolves with
redshift.  
Comparing the measurements in the $z=[2.2,2.5]$ and $z=[3.,3.5]$
intervals, we find a $50 \pm 10\%$ decrease in the lowest SDSS
redshift bin.
This evolution occurs during the same time interval when 
the decrease in $\ldla(X)$ is observed (Figure~\ref{fig:taux}).
Regarding the line density, 
we have attributed the decline to a decrease in the typical
cross-section of the damped \lya\ systems.  
Given the decrease in \odla\, it is evident that the gas 
is not simply contracting into smaller structures (e.g.\ via dissipative
cooling), but that the gas is being consumed and/or expelled from
the system.

Contrary to previous claims, we find no evidence for  
evolution in \odla\ at $z>3.5$ although a modest increase (or decrease)
is permitted by the observations.  Similarly, the results at $z \approx 2$
from previous work are consistent (albeit uncomfortably 
high\footnote{
Note that the value for \odla\ is reduced by $\approx 50\%$
if one removes the single damped system at $z=2.04$ toward Q0458$-$02.
This stresses the sensitivity of these low redshift results
to small number statistics (as stressed by Chen et al.\ 2005)
and the bootstrap error estimate may be overly optimistic in this case.
Coincidentally, Q0458$-$02 is an optically variable quasar
which in its low state has a magnitude below the original survey detection
limit.  Also note that 
this redshift interval does not include any data
from the SDSS survey.}) 
with the $z = 2.3$ data point.
We also caution the reader
that the results at $z>4$ should be confirmed by
higher resolution observations.
At these redshifts, blending in the \lya\ forest is severe and we worry
that the SDSS spectroscopic resolution is insufficient.
In passing, we also note that a significant contribution to
\odla\ at $z>4$ comes along the single sightline to 
J162626.5+275132 which shows two $\mnhi \approx 10^{21} \cm{-2}$
damped \lya\ systems at $z \approx 5$ and even two additional systems with
comparable \nhi\ value at $z \sim 4.5$ which is not in the 
statistical sample.

\begin{sidewaystable*}\footnotesize
\begin{center}
\caption{SUMMARY\label{tab:summ}}
\begin{tabular}{lccccccccccc}
\tableline
\tableline
$z$ & $dX$ & $m_{\rm DLA}$ & $\bar z^a$ & & $\log f_{\rm HI}(N,X)$
& & & $\ell_{\rm DLA}(X)$ & $\Omega_g^{\rm DLA}$ & $\alpha_{\rm LLS}$ & 
$\Omega_g^{\rm LLS}$ \\
& & & & $\mnhi \epsilon [20.3, 20.6)$ & 
$\mnhi \epsilon [20.6, 21.0)$ & 
$\mnhi \epsilon [21.0, 21.4)$ & 
$\mnhi \epsilon [21.4, 21.8)$ & & ($\times 10^{-3}$)
& & ($\times 10^{-3}$) \\
\tableline
$\lbrack$2.2,5.5]$^b$ & 7333.2& 525&3.06&$-21.71^{+0.03}_{-0.03}$&$-22.45^{+0.04}_{-0.03}$&$-23.21^{+0.05}_{-0.05}$&$-24.14^{+0.10}_{-0.09}$&$0.072^{+0.003}_{-0.003}$&$0.817^{+0.050}_{-0.052}$&$-1.0$&0.26\\
$\lbrack$1.7,2.2]& 420.7&  30&1.94&$-21.88^{+0.14}_{-0.12}$&$-22.26^{+0.12}_{-0.11}$&$-23.20^{+0.23}_{-0.19}$&$-24.20^{+0.45}_{-0.32}$&$0.071^{+0.015}_{-0.011}$&$0.842^{+0.228}_{-0.252}$\\
$\lbrack$2.2,2.5]&1796.4&  95&2.36&$-21.84^{+0.06}_{-0.06}$&$-22.52^{+0.08}_{-0.07}$&$-23.59^{+0.17}_{-0.15}$&$-24.36^{+0.26}_{-0.22}$&$0.053^{+0.006}_{-0.005}$&$0.501^{+0.071}_{-0.068}$&$-0.9$&0.21\\
$\lbrack$2.5,3.0]&2984.7& 205&2.75&$-21.73^{+0.04}_{-0.04}$&$-22.47^{+0.06}_{-0.05}$&$-23.26^{+0.09}_{-0.08}$&$-24.10^{+0.15}_{-0.13}$&$0.069^{+0.005}_{-0.005}$&$0.849^{+0.084}_{-0.090}$&$-1.0$&0.24\\
$\lbrack$3.0,3.5]&2068.7& 169&3.22&$-21.72^{+0.05}_{-0.05}$&$-22.36^{+0.06}_{-0.06}$&$-23.02^{+0.08}_{-0.08}$&$-24.12^{+0.18}_{-0.16}$&$0.082^{+0.007}_{-0.006}$&$1.036^{+0.105}_{-0.108}$&$-1.0$&0.27\\
$\lbrack$3.5,4.0]&1066.5&  89&3.72&$-21.65^{+0.06}_{-0.06}$&$-22.47^{+0.10}_{-0.09}$&$-23.03^{+0.12}_{-0.11}$&$-23.91^{+0.20}_{-0.18}$&$0.083^{+0.009}_{-0.008}$&$1.076^{+0.168}_{-0.173}$&$-1.1$&0.28\\
$\lbrack$4.0,5.5]& 426.1&  42&4.35&$-21.51^{+0.09}_{-0.08}$&$-22.37^{+0.14}_{-0.12}$&$-23.21^{+0.23}_{-0.19}$&$-24.21^{+0.45}_{-0.32}$&$0.099^{+0.017}_{-0.014}$&$0.974^{+0.215}_{-0.188}$&$-1.0$&0.45\\
$\lbrack$3.5,5.5]&1492.6& 131&3.92&$-21.61^{+0.05}_{-0.05}$&$-22.43^{+0.08}_{-0.07}$&$-23.07^{+0.10}_{-0.10}$&$-23.98^{+0.18}_{-0.16}$&$0.088^{+0.008}_{-0.007}$&$1.047^{+0.127}_{-0.137}$&$-1.1$&0.32\\
\tableline
\end{tabular}
\end{center}
\tablenotetext{a}{Mean absorption redshift}
\tablenotetext{b}{Restricted to SDSS-DR3\_4.}
\end{sidewaystable*}

\subsection{Summary}

Table~\ref{tab:summ} presents a summary of \fnhi\ and its zeroth
and first moments for a series of redshift intervals.   
A qualitative summary of the new results is that
they are in broad agreement with the previous analyses
based on substantially smaller datasets \citep{wolfe95,storrie00,peroux03}.
With the increased sample size, however, we have measured evolution
in the incidence and \ion{H}{1} content of these galaxies at $z>2$
to unprecedented precision.

\clearpage

\section{SYSTEMATIC EFFECTS}
\label{apx:sys}

With the large sample size of the SDSS-DR3, we now have sufficient
signal to examine a number of systematic effects on the
results of our analysis.  This section describes an investigation into
a few of these effects.  It is advantageous that the
entire SDSS sample
was observed with the same instrument, reduced with
the same software, and analyzed by one group.  
In contrast, the compilation of \cite{peroux03}
is based on many surveys carried out on over 10 telescopes
and reduced and analyzed by at least five different individuals.
This undoubtedly leads to at least minor systematic differences
in the surveys, e.g.\ differences in the continuum placement,
systematic effects related to instrumental resolution, and/or color
selection of the quasar populations.
Of course, if multiple groups report the same value, this lends
confidence to the sets of procedures related to damped \lya\ surveys.

\begin{figure}[ht]
\begin{center}
\includegraphics[width=3.6in]{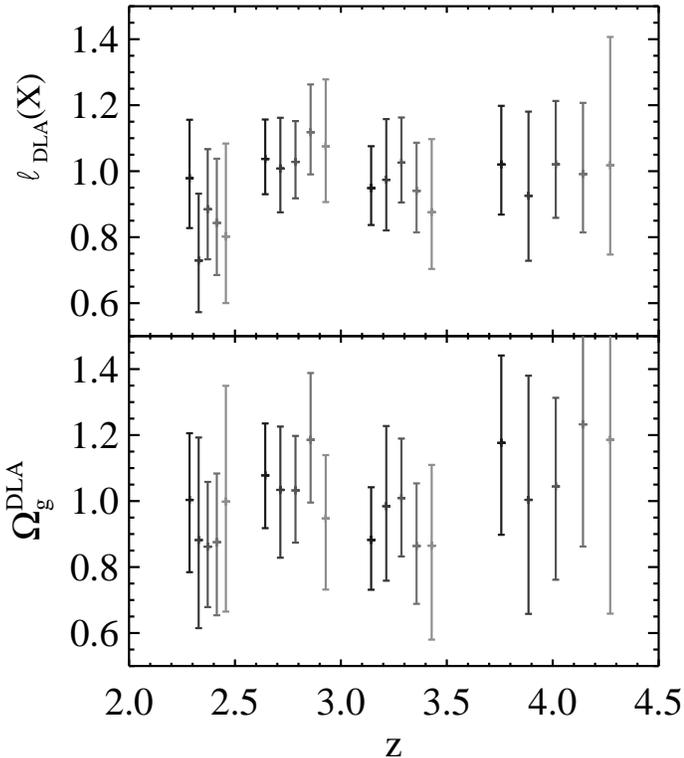}
\caption{The top panel shows the line density as a function of redshift
for the various sample cuts on \snrlim\ and color selection criteria
listed in Table~\ref{tab:cuts}.  The data are plotted relative to the
SDSS-DR3\_4 sample and are offset in redshift for presentation purposes
only.  The values are calculated in the redshift intervals:
$z=[2.2, 2.5], z=[2.5,3.], z=[3.,3.5], z=[3.5,4.4]$.
The lower panel shows the results for \odla\ for the same cuts and also
relative to SDSS-DR3\_4.
One notes very little difference between the various samples.
}
\label{fig:statsmpl}
\end{center}
\end{figure}

\subsection{\snrlim\ and Color Selection}

All of the results presented in this paper are based on searches
for damped \lya\ systems in the spectral regions of quasars where
the median SNR exceeds 4 in a 20~pixel bin ($\S$~\ref{sec:redpath}).
Our experience via visual inspection and Monte Carlo simulations
is that this limit is just
satisfactory.  Lowering the limit to a SNR$_{lim} = 3$
gives many more false positives
and a greater number of missed damped \lya\ systems.
Of course, it is important to test the sensitivity of our results
to \snrlim.
Figure~\ref{fig:statsmpl} presents the (a) line density $\ldla(X)$
and (b) neutral gas mass density \omg\ 
relative to the full sample as a function of redshift
for several values of \snrlim\ or the 
samples restricted to the strict color criteria 
\citep{richards02}.  See Table~\ref{tab:cuts} for a description
of each cut of the SDSS-DR3.

Examining the figure, we find little difference between the
various samples.  We find, therefore, no 
systematic bias related to \snrlim\ or the color-selection
criteria.  In fact, the scatter in the data is smaller than that
expected
for a Poissonian distribution.  This is because the samples are
not independent.  Nevertheless, this analysis demonstrates that
our results are relatively insensitive to 
\snrlim\ and the color selection criteria.

\begin{figure}[ht]
\begin{center}
\includegraphics[height=3.6in,angle=90]{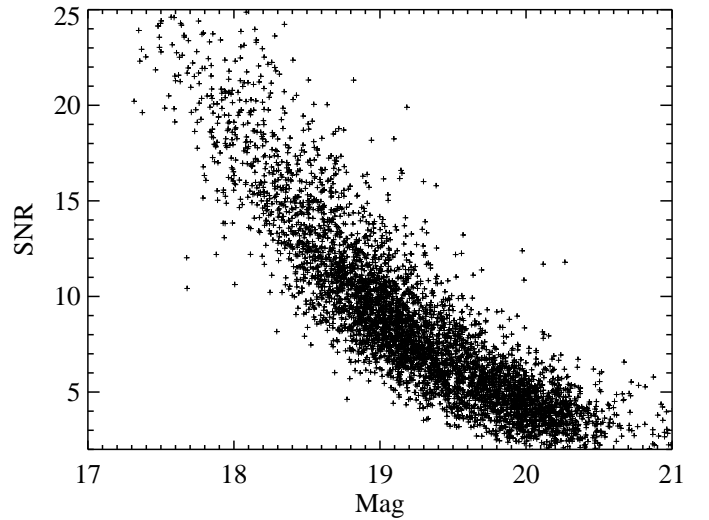}
\caption{
Plot of the quasar magnitude at $\lambda_{rest} = 1300$ to 1350\AA\
against the median spectral SNR at $\lambda \approx 1400$\AA.
There is a good correlation between the two quantities although
there are many outliers due primarily to BAL quasars.
}
\label{fig:magsnr}
\end{center}
\end{figure}

\subsection{Quasar magnitude and SNR}
\label{sec:qsomag}

There are a number of reasons to examine the results as a function
of the quasar magnitude and/or the SNR of its spectrum.
These include considerations of dust obscuration \citep{fall93},
the precision of the \nhi\ values as a function of SNR,
and gravitational lensing.
We consider the quasar magnitude and spectral SNR together because
we have found a similar dependence on the results for these
two characteristics.  Although there is not a strict one-to-one
relationship between quasar magnitude and SNR for the SDSS survey,
the two characteristics are closely associated.  

We have calculated the magnitude of the quasar as follows:
First, we identify the Sloan filter closest to $\lambda = (1+z_{qso}) 1325$\AA.
Second, we adopt the PSF magnitude for the quasar as reported in 
SDSS-DR3.  Third, we flux calibrate the 1D quasar spectrum assuming
its relative flux is accurate.
Finally, we calculate the median flux in the interval
$\lambda_{rest} = 1300$ to 1350\AA\
(which avoids strong emission line features) and convert to AB magnitude.
To estimate the SNR of the spectrum, we calculate the median SNR for
$\lambda_{rest} = 1440$ to 1490\AA\ where $z_{qso} < 5$ and
$\lambda_{rest} = 1287$ to 1366\AA\ for $z_{qso} > 5$.
Figure~\ref{fig:magsnr} plots the quasar magnitude versus the
spectral SNR.  There is a reasonably tight correlation with a few
outliers which are primarily exotic BAL quasars.  

\begin{figure}[ht]
\begin{center}
\includegraphics[width=3.6in]{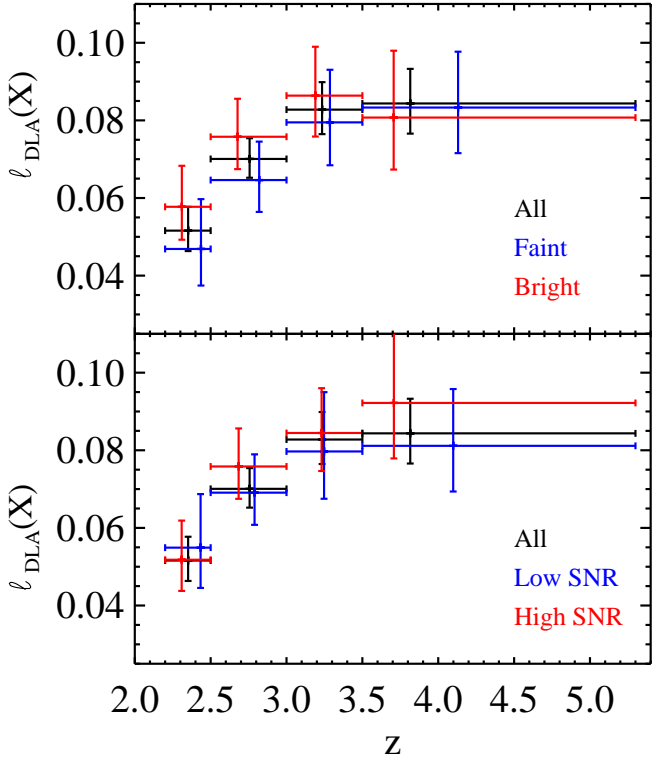}
\caption{
Line density as a function of redshift for the full sample, the
brightest 33$\%$ of the quasars and the faintest 33$\%$ of the
quasars in each redshift interval.  There is a mild trend toward
higher line density toward the brighter quasars, but the effect
is not statistically significant.  Similar results are seen
for the quasars cut on spectral SNR.
}
\label{fig:lxmag}
\end{center}
\end{figure}

The upper panel in 
Figure~\ref{fig:lxmag} presents the line density of damped 
\lya\ systems as a function of redshift for three cuts of the
SDSS-DR3\_4 sample: 
(i) the complete sample; 
(ii) the brightest 33$\%$ of the quasars; and
(iii) the faintest 33$\%$ of the quasars.
One notes no significant dependence of $\ldla$ on quasar magnitude.
The lower panel shows the same quantity for cuts on the SNR.
Note that the median magnitude of the bright/faint sample 
for the lowest redshift bin is 18.4 and 19.5\,mag respectively.
This difference in magnitudes is significant and it nearly brackets the 
`break magnitude' of the quasar luminosity function at $z=2$ \citep{boyle00}.
The results indicate that if dust obscuration is relevant to damped
\lya\ surveys, then the line density is insensitive to its effects.
While the line density may not be very sensitive to dust obscuration,
it is possible that our results violate the predictions of
\cite{fall93}.  A quantitative statement awaits a full treatment
which includes the SDSS quasar color selection criteria
(Murphy et al., in prep).

\begin{figure}[ht]
\begin{center}
\includegraphics[width=3.6in]{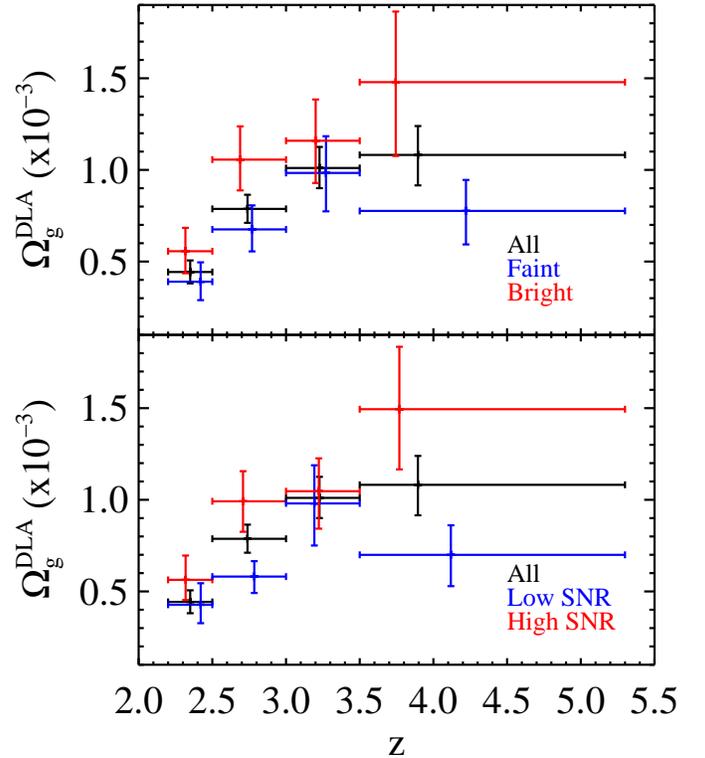}
\caption{
Gas mass density as a function of redshift for the full sample, the
brightest 33$\%$ of the quasars and the faintest 33$\%$ of the
quasars in each redshift interval.  There is a systematic trend of
higher \odla\ values toward the brighter quasars.  Although the effect
is not significant at the $2\sigma$ level in any given redshift bin,
the effect is significant at $> 95\%$ c.l. when considering all bins together.
In the text we discuss a number of possible explanations, the most likely 
being gravitational lensing.
}
\label{fig:omgmag}
\end{center}
\end{figure}

\begin{figure}[ht]
\begin{center}
\includegraphics[height=3.6in,angle=90]{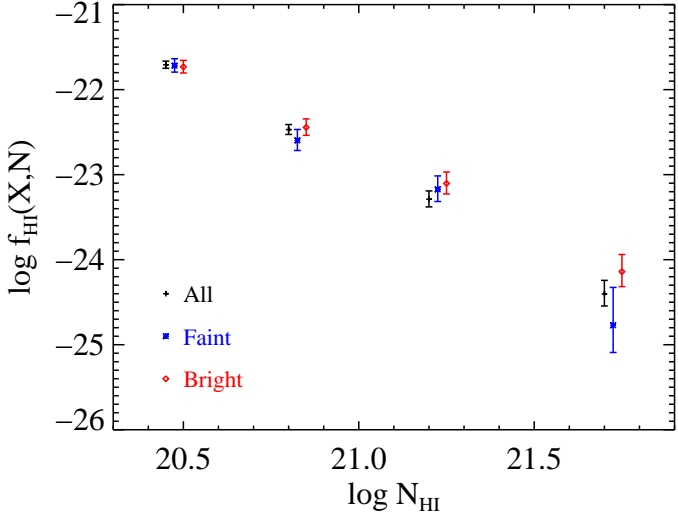}
\caption{
The \ion{H}{1} column density distribution at $z=2.5$ to 3
for the full sample, the
brightest 33$\%$ of the quasars and the faintest 33$\%$. 
One observes no statistically significant difference in 
\fnhi\ at low \nhi\ value but that the bright sample has 5$\times$
higher value at large \nhi.
}
\label{fig:fnmag}
\end{center}
\end{figure}

In contrast, we do find a systematic relationship between quasar
magnitude (and SNR) with \odla.  Figure~\ref{fig:omgmag} presents the
results for the same cuts of SDSS-DR3\_4 as in Figure~\ref{fig:lxmag}.
It is evident that the bright sub-sample shows systematically higher
\odla\ values than the faint sub-sample.  Of course, in order to explain
the independence of $\ldla$ to quasar magnitude, all of the difference
in \odla\ must arise from the frequency of large \nhi\ damped systems.
Indeed, Figure~\ref{fig:fnmag} shows that the \fnhi\ distributions
are comparable for the bright and faint sub-samples for 
$\mnhi < 10^{21} \cm{-2}$ but there is a factor of 5 difference in the
incidence of systems with $\mnhi > 10^{21.4} \cm{-2}$.   This
has the greatest impact on the \odla\ results which we now discuss.

Let us consider possible explanations for the results in 
Figures~\ref{fig:omgmag} and \ref{fig:fnmag}:

{\bf 1. Chance (small number statistics):}
In any given redshift interval in Figure~\ref{fig:omgmag} the results
are significant at only the $\approx 1 \sigma$ level.  Put together,
however, the effect is significant at greater than the $95\%$ c.l.;
the weighted mean of the ratio of the bright to faint sub-samples
is 
$\Omega_g^{\rm DLA}({\rm bright})/\Omega_g^{\rm DLA}({\rm faint}) = 
1.4 \pm 0.2$.
Therefore, we consider it unlikely that the effect is related only to 
statistical fluctuations.

\begin{figure}[ht]
\begin{center}
\includegraphics[height=3.6in,angle=90]{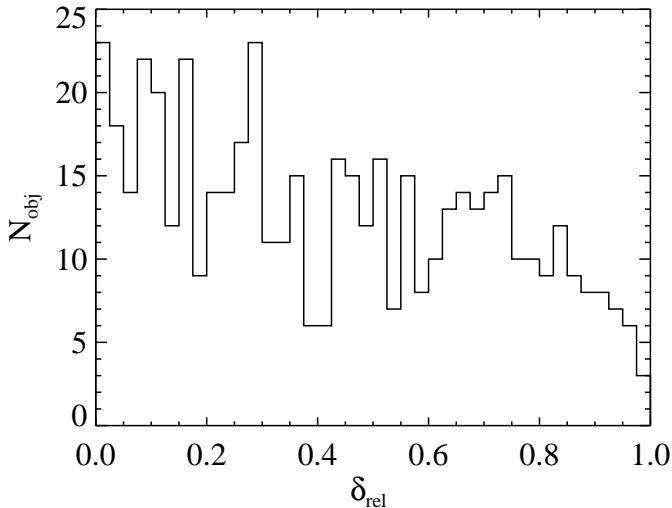}
\caption{
Histogram of the relative separation of the damped \lya\ absorption
redshift from the redshifts defining the survey path of the
background quasar ($z_f$ and $z_i$) as defined by 
Equation~\ref{eqn:relsep}.
One notes fewer damped \lya\ systems with $z \approx z_i$ as
expected because the incidence of damped \lya\ systems increases
with redshift.
}
\label{fig:sep}
\end{center}
\end{figure}

{\bf 2. DLA are intrinsic to the quasar:}
This is a difficult assertion to disprove. 
Figure~\ref{fig:sep} plots a histogram of the relative statistic
of the damped \lya\ systems from the quasar

\begin{equation}
\delta_{rel} \equiv \frac{z_f - z_{DLA}}{z_f - z_i}
\label{eqn:relsep}
\end{equation}

\noindent restricted to sightlines where $z_f - z_i > 0.2$.
While there are fewer systems with $z_{DLA} \approx z_i$, this is
expected because the incidence of the damped
\lya\ systems increases with redshift.  At present, we do not consider 
association with the quasar to be an important issue.

{\bf 3. The survey has missed damped \lya\ systems with large
\nhi\ toward faint quasars:}
Given that our algorithm is most sensitive to spectral regions
of very low SNR, it is very unlikely that we would miss a damped 
\lya\ system whose core is much larger than the search window $6(1+z)$\AA.
Furthermore, the profiles with very large \nhi\ value are the easiest
to identify and we would have recovered most in our 
visual inspection of the spectra.  Finally, these systems will probably all
show significant metal-line absorption and we would have identified
them independently of the \lya\ profile (e.g.\ $\S$~\ref{sec:bump}).

There is, however, one bias related to this point.  If a damped
\lya\ profile coincides with the wavelength $\lambda_i$
in the quasar
spectrum where the median SNR would have otherwise exceeded \snrlim\
(see $\S$~\ref{sec:redpath}), then the presence of a damped \lya\
system would preclude its inclusion in the statistical sample
(and possibly its discovery altogether).
That is, the presence of a damped \lya\ profile at wavelength
$\lambda_{DLA}$, 
where the median SNR of the unabsorbed flux exceeds 4, 
lowers the SNR at $\lambda_{DLA}$
such that $\lambda_i$ occurs redward of the damped \lya\ system.
This is true for damped \lya\ systems of all \nhi\ values,
although the effect will be greatest for larger \nhi.
We have accounted for this bias, in part, by defining $z_1$ to
be 1000\,\kms\ redward of the redshift corresponding to $\lambda_i$.
Nevertheless, this issue is a concern, albeit one we consider minor.

{\bf 4. Overestimating \nhi\ in the bright sample:}
We have reexamined the damped \lya\ profiles for every system with
$\mnhi > 10^{21.4} \cm{-2}$.  In no case would we modify the best
\nhi\ value beyond the reported uncertainty.  Furthermore,  the
bright sample has the highest SNR data and we consider the \nhi\
values to be well constrained.  All have associated metal-line absorption
and reasonably well determined quasar continua.  Therefore, we consider
this explanation to be unimportant.

\begin{figure*}[ht]
\begin{center}
\includegraphics[height=3.4in,angle=90]{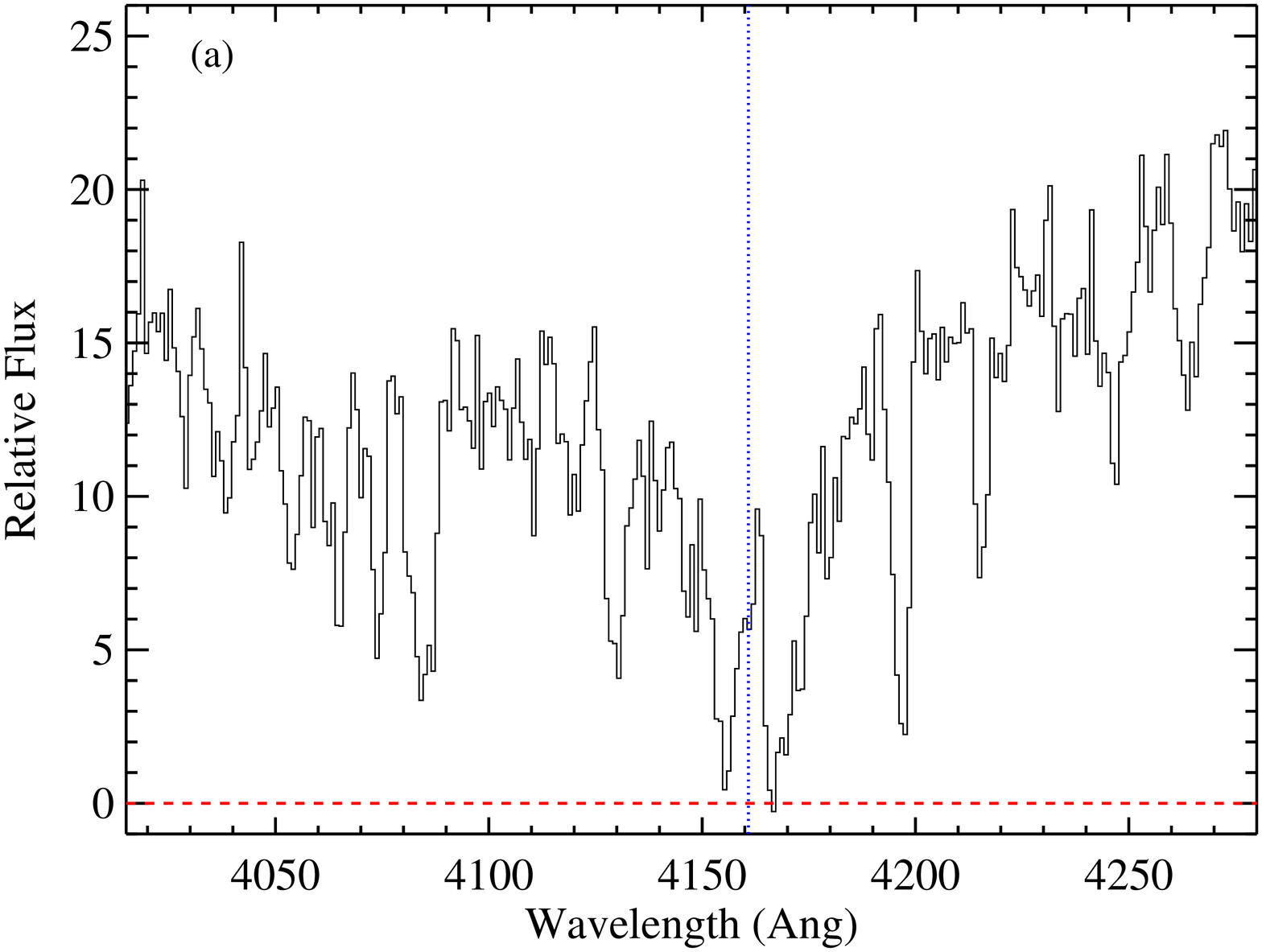}
\includegraphics[width=3.6in]{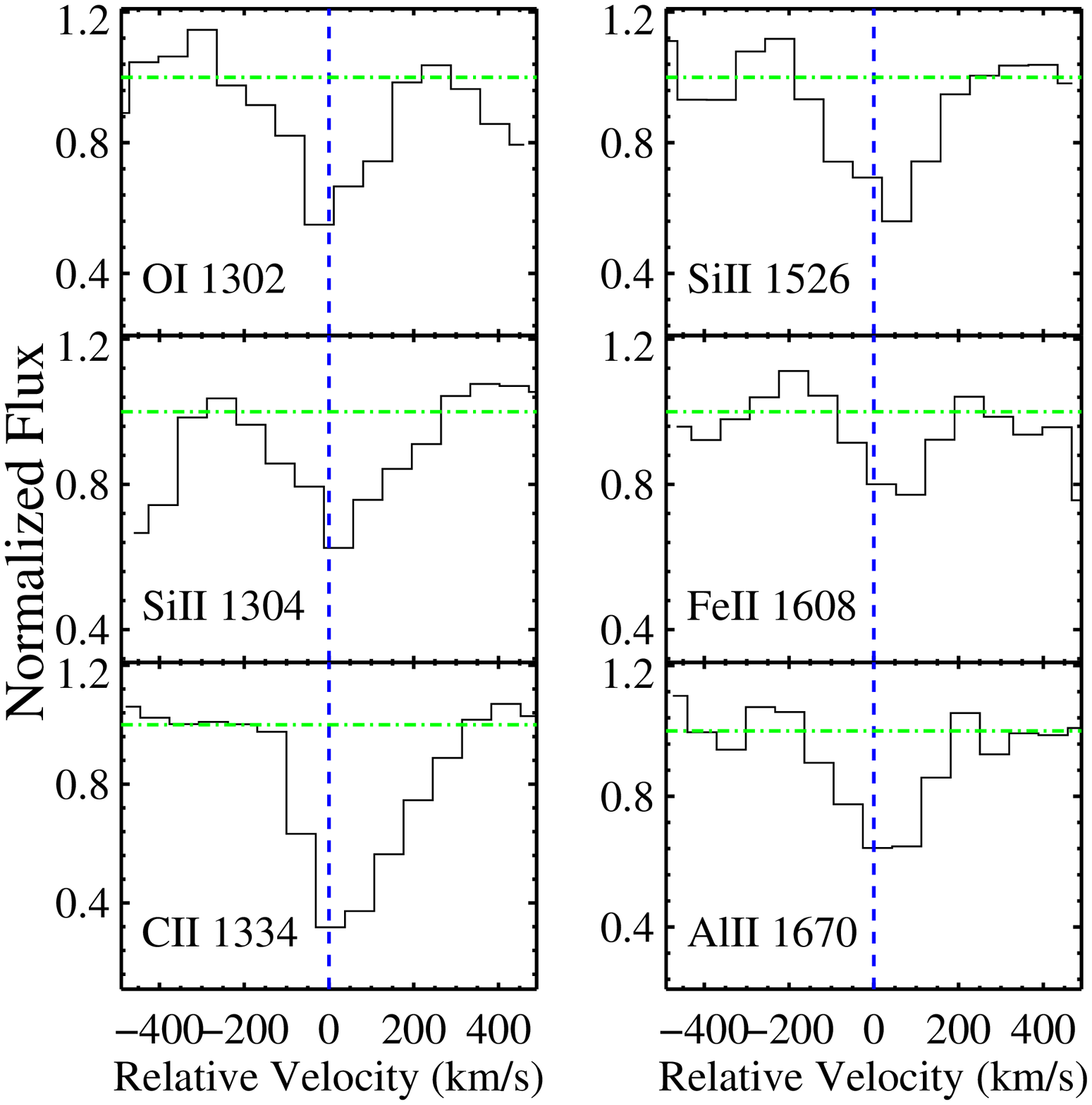}
\caption{
The (a) \lya\ profile and (b) metal-line profiles of the super-LLS
at $z=2.42$ toward J130634.6+523250.  The emission-line feature at
the center of the \lya\ profile is due to [OII] emission from a
galaxy at $z=0.112$ in the SDSS fiber.  If this absorber had no
significant metal-line absorption, it would have been dropped
from our sample altogether.
}
\label{fig:lyaemiss}
\end{center}
\end{figure*}

{\bf 5. Underestimating \nhi\ in the faint sample:}
Because these quasars are faint, the spectra have lower
SNR and more uncertain \nhi\ values.  While this could lead to 
a significant underestimate of \nhi, several effects work against it.
First, \fnhi\ is sufficiently steep (particularly at high \nhi)
that a Malmquist bias will actually lead to a flattening of \fnhi\
at the 
high \nhi\ end.  This leads to an overestimate of \odla\ in the 
faint subset because the
bias will be more significant for \nhi\
values with larger uncertainty. 
Second, the effects of low SNR are most significant at low \nhi\
values where only a handful of pixels may constrain the result.
But these systems do not contribute significantly to \odla.
Third, at large \nhi\ value, the profile extends over $\approx 100$\AA\
and is generally constrained by tens of pixels.  It is true that
uncertainty in the continuum level is greater, but we do not
believe this to be the problem.

To further pursue the possibility that we have underestimated
\nhi\ in the faint sub-sample, we reexamined every \lya\ profile
with $\mnhi > 10^{21} \cm{-2}$ in the $z \sim 2.7$ redshift bin.  
While we could allow for 0.1 to 0.2\,dex larger values in a number
of cases, this would not account for the differences in \odla.
Furthermore, we note that the \fnhi\ value in the \nhi\ interval
$10^{21} - 10^{21.4} \cm{-2}$
is lower in the faint sub-sample.
Therefore, it would require a systematic underestimate in
\nhi\ at nearly all values to explain the observations.  At present,
we consider an underestimate of \nhi\ values in lower SNR data
to be a possible contribution to the differences in \odla\
but unlikely to be the main explanation.

{\bf 6.  Dust Obscuration:}
To date, every damped \lya\ system observed at high resolution
has a measured metallicity exceeding 1/1000 solar \citep{pro03b}.
With the presence of metals, there is the prospect that the gas
has a non-negligible dust content \citep[e.g.][]{pettini94}.
Indeed, damped \lya\ systems with metallicity $> 1/10$ solar
have enhanced Si/Fe ratios which are best explained by differential
depletion \citep{pro03}.  
As \cite{ostriker84} first noted, the damped \lya\ systems with the
highest optical depths of dust will obscure the background quasar
and possibly remove it from a magnitude limited quasar survey.

The prediction, as formalized by \cite{fall93}, is that optical
quasar samples will underestimate \odla\ owing to this bias.
As \cite{ellison04} have stressed, the effect will be less
significant if the quasar survey extends beyond the peak in the
luminosity function.   In any case, this
bias implies that the \odla\ values from brighter quasars should be
lower than the value inferred from faint quasars.
We observe the opposite trend in Figure~\ref{fig:omgmag}.
Therefore, the results are unlikely to be explained by dust
obscuration.

{\bf 7. Gravitational Lensing:}
If the damped \lya\ systems arise in massive halos or disks, then
it is possible that they would gravitationally magnify
the background quasar \citep{bart96,smette97,maller97}.
In their analysis of the SDSS-DR2 quasar database,
\cite{murphy04} reported an approximately $2\sigma$
result that the luminosity function of quasars with foreground
damped \lya\ systems is brighter than those without.
Unlike dust obscuration, this systematic effect could explain
the results in Figure~\ref{fig:omgmag}.  In fact, treatments
based on describing damped \lya\ systems as exponential
disks predict that the effect will be greatest for damped 
\lya\ systems with large \nhi\ value \citep{bart96,maller97}.

At present, we believe gravitational lensing to be the most viable
explanation for the results in Figures~\ref{fig:omgmag} 
and \ref{fig:fnmag}.
The key implications for \odla\ are that 
(i) our reported values may be too high
and
(ii) the evolution of \odla\ revealed by the entire sample is
qualitatively correct.
Therefore, while this systematic effect is important, 
our conclusions are relatively invariant to it.
We note, however, that the effect should be even more pronounced at
$z<2$ and may significantly affect damped \lya\ statistics at
these redshifts \citep{rao00}.

If gravitational lensing does explain the effect, it may
allow for a statistical mass measurement of at least the high
\nhi\ sightlines.
It also motivates a
high spatial resolution survey of all damped \lya\ systems
with $\mnhi > 10^{21} \cm{-2}$.  In a future paper
(Murphy et al.\ 2005, in prep.) we will present a full analysis 
of gravitational
lensing and dust obscuration for the SDSS-DR3 sample.

\subsection{Human Error}

There is a non-negligible likelihood that we have overlooked
a few systems or have made significant errors in a few select systems.
Regarding PH04, for example, we note the incorrect \nhi\ value
for the damped system at $z=2.77$ toward J084407.29+515311
($\S$~\ref{sec:acc}) and also a bug in our calculation of $\ell(z)$
for the SDSS-DR1 sample.
To quickly disseminate corrected and updated results to the
community, we have established a public web site where all of the
fits and analysis will be presented
(http://www.ucolick.org/$\sim$xavier/SDSSDLA/index.html).
We encourage the community to report any mistakes with our
analysis to the lead author via the email address
sdssdla@ucolick.org.

\subsection{Unusual Systematic Effects: ``Things that go bump in the night''}
\label{sec:bump}

With a sample of damped \lya\ systems approaching 1000, it
is not surprising that unexpected systematic errors will arise.
Figure~\ref{fig:lyaemiss} presents the \lya\ profile and 
metal-line profiles for the damped \lya\ candidate at $z=2.42$
toward J130634.6+523250.  This damped \lya\ candidate was
not identified by our automated algorithm because of 
significant flux at the center of the \lya\ profile.
Instead, the system was identified because of its metal-line
absorption and we immediately hypothesized that the
flux in the \lya\ profile was due to \lya\ emission from the host 
galaxy.  The emission line would be amazingly strong, however,
and we considered alternate explanations.  In due time, we
realized that the feature is an emission line: [OII] emission
from a $z=0.116$ galaxy which lies within the $3''$ SDSS fiber.
Emission lines of H$\alpha$, H$\beta$, and [OIII] are also 
apparent in the quasar spectrum.  Ignoring the [OII] emission,
we have fit the \lya\ profile and its central value places
it beneath the statistical threshold for damped \lya\
systems.  Nevertheless, this is a systematic effect   
which leads to an underestimate of \fnhi\ at all \nhi\ value.
It is difficult to quantify the overall effect here, but it
is presumably less than $1\%$.

\section{DISCUSSION AND SPECULATIONS}
\label{sec:discuss}

The emphasis of this paper is to describe the results of the
damped \lya\ survey of the SDSS-DR3 quasar database.  These results
and a discussion of the systematic errors
were presented in the previous sections. 
We now consider a few
of the implications with emphasis on the new results.  We also
compare the observations against theoretical treatments of the damped 
\lya\ systems within $\Lambda$CDM models of galaxy formation.
We consider the results from the smooth particle hydrodynamic (SPH)
simulations of \cite{nagamine04a}, the Eulerian simulations of 
\cite{cen03}, and the semi-analytic model (SAM) of the Santa Cruz 
group \citep{spf01,maller01,maller03}.
It is important to stress that each model includes its own set
of star formation and feedback recipes which do bear on the results
for the damped \lya\ systems.

Consider first the \nhi\ frequency distribution, \fnhi.
Perhaps the most remarkable result from the SDSS-DR3 sample is
that there is no statistical evidence for any evolution in the
shape of \fnhi\ with redshift (Figure~\ref{fig:fncomp}).
There is, however, evidence for evolution in the 
normalization of \fnhi\ as traced by the trends in the zeroth
and first moments of the distribution function. 
These results suggest that the gas distribution within galaxies
is similar at all redshifts and that only the number and/or sizes
of these galaxies evolve significantly.
Another interesting result is that the faint-end slope of the
\fnhi\ distribution is $\alpha_3 \approx -1.8$.  This slope
matches the faint-end slope of the dark matter halo mass function
for CDM \citep[e.g.][]{sheth01}.
If this is not a coincidence, it indicates that low mass halos
dominate the incidence of damped \lya\ systems at low \nhi\ values.
Furthermore, it suggests that the cross-section $A(X)$ of
low mass galaxies is nearly independent of mass.
At present, however, we consider the correspondence to be a coincidence.

\begin{figure}[ht]
\begin{center}
\includegraphics[width=3.6in]{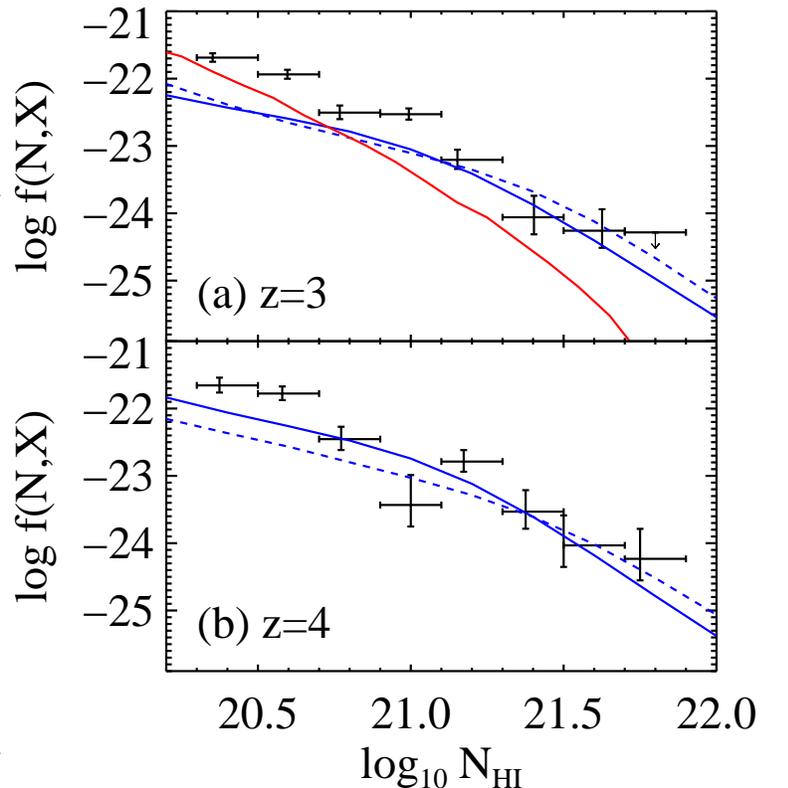}
\caption{
The \ion{H}{1} frequency distribution of the damped \lya\
systems at $z=3$ (top; specifically $z=2.8$ to 3.2) and 
$z=4$ (bottom; specifically $z=3.7$ to 4.3) compared against
the theoretical curves of \cite{maller01} (SAMS; lighter curve) and
\cite{nagamine04a} (SPH; darker curves; dashed is the D5 model and
solid is the Q5 model).
}
\label{fig:nagfn}
\end{center}
\end{figure}

A comparison of the results against $\Lambda$CDM models of galaxy
formation is presented
in Figure~\ref{fig:nagfn} at $z=3$  and
$z=4$. 
The \fnhi\ curves are for the SPH simulations of
\cite{nagamine04a} and the SAM model of \cite{maller01}.  
The SAM model shows a reasonable match to the shape of \fnhi\
at $z=3$, yet systematically underpredicts the observations.  
\cite{maller01} were primarily interested in modeling the kinematics
of the damped \lya\ systems and we suspect their results are insensitive
to the normalization of \fnhi.  Nevertheless, the discrepancy suggests
that the SAM model has too few damped \lya\ systems at $z=3$.
Similarly, the SPH theoretical curves offer a reasonable
match to the observations at large \nhi\ value yet significantly 
underpredict \fnhi\ at low \nhi\ values.  

The discrepancy has several
important implications.  First, the SPH simulations will predict
a significantly smaller contribution to \ohi\ for the Lyman limit
systems at these redshifts.  In part, this is a restatement
of a current problem in numerical cosmology: the simulations
underpredict the incidence of Lyman limit systems by an order
of magnitude \citep{gardner01}.
Second, we reemphasize the critical constraint imposed by
two sets of observations of the damped \lya\ systems: (a) \fnhi\
and (b) the velocity width distribution of the low-ion profiles
\citep{pw97}.  \cite{jdzk98} first emphasized that 
 there is a tension between \fnhi\ 
and the velocity width distribution.  
Specifically, if one introduces enough low mass halos
to match the low \nhi\ end of \fnhi\ this implies far too
many damped \lya\ systems with small velocity width.
Therefore, if the SPH simulations were modified to match the
\fnhi\ observations (e.g.\ via different treatments of feedback
or radiative transfer), we predict that (i) the dependence of the
cross-section $A(X)$ on the halo mass as
reported by \cite{nagamine04a} would change 
significantly and (ii) the simulations would not reproduce the
damped \lya\ kinematics.
We also note that the results of \cite{nagamine04a} show that
\fnhi\ is relatively insensitive to treatments of feedback.  
It is possible that \fnhi\ and the damped 
\lya\ kinematics present a fundamental challenge to scenarios
of galaxy formation within the $\Lambda$CDM cosmology.

Now consider the evolution in the line density of damped \lya\
systems.
Because $\ldla(X)$ is the product of the comoving number density and
cross-section of damped \lya\ systems (Equation~\ref{eqn:cover}),
the rise in $\ldla(X)$ reflects an increase in one or both of
these quantities.  
We can estimate the evolution in $n_{\rm DLA}$ by considering
the Press-Schechter formalism for the mass function of
dark matter halos \citep[e.g.][]{peacock99}.  Within this formalism 
one can define a mass scale $M_*$ as
a function of redshift which identifies the typical mass of assembly.
At $z=3$ in the $\Lambda$CDM cosmology with $\sigma_8 =1$, 
$M_* \approx 10^{10}$\,M$_\odot$.  Halos with masses $M \gg M_*$
will have a number density which increases significantly from $z=3$ to 2.
Therefore, damped \lya\ systems are unlikely to arise in halos with
$M \gg 10^{10}$\,M$_\odot$ at these redshifts.
Furthermore, although halos with $M < M_*$ do have a decreasing
co-moving number density, the decrease is small
(i.e. $<15\%$ for $M=10^9$\,M$_\odot$).  Within the context of the
hierarchical cosmology, therefore, the most likely explanation for the
decrease in $\ldla(X)$ is a corresponding decline in $A(X)$.
There are several physical effects which could reduce $A(X)$, e.g.,
star formation, AGN feedback, photoionization, and galactic winds.
Current estimates of the star formation rates are relatively uncertain
over this redshift range \citep[e.g.][]{wgp03} but probably
cannot account for the bulk of evolution.
Similarly, the intensity of the
extragalactic background radiation field is believed to be
roughly constant \citep[e.g.][]{haardt96} and photoionization effects
should be minor.  Therefore, we contend that one or more feedback
mechanisms have significantly reduced the typical cross-section of
galaxies to damped \lya\ absorption from $z=3$ to 2.

\begin{figure*}[ht]
\begin{center}
\includegraphics[height=5.5in,angle=90]{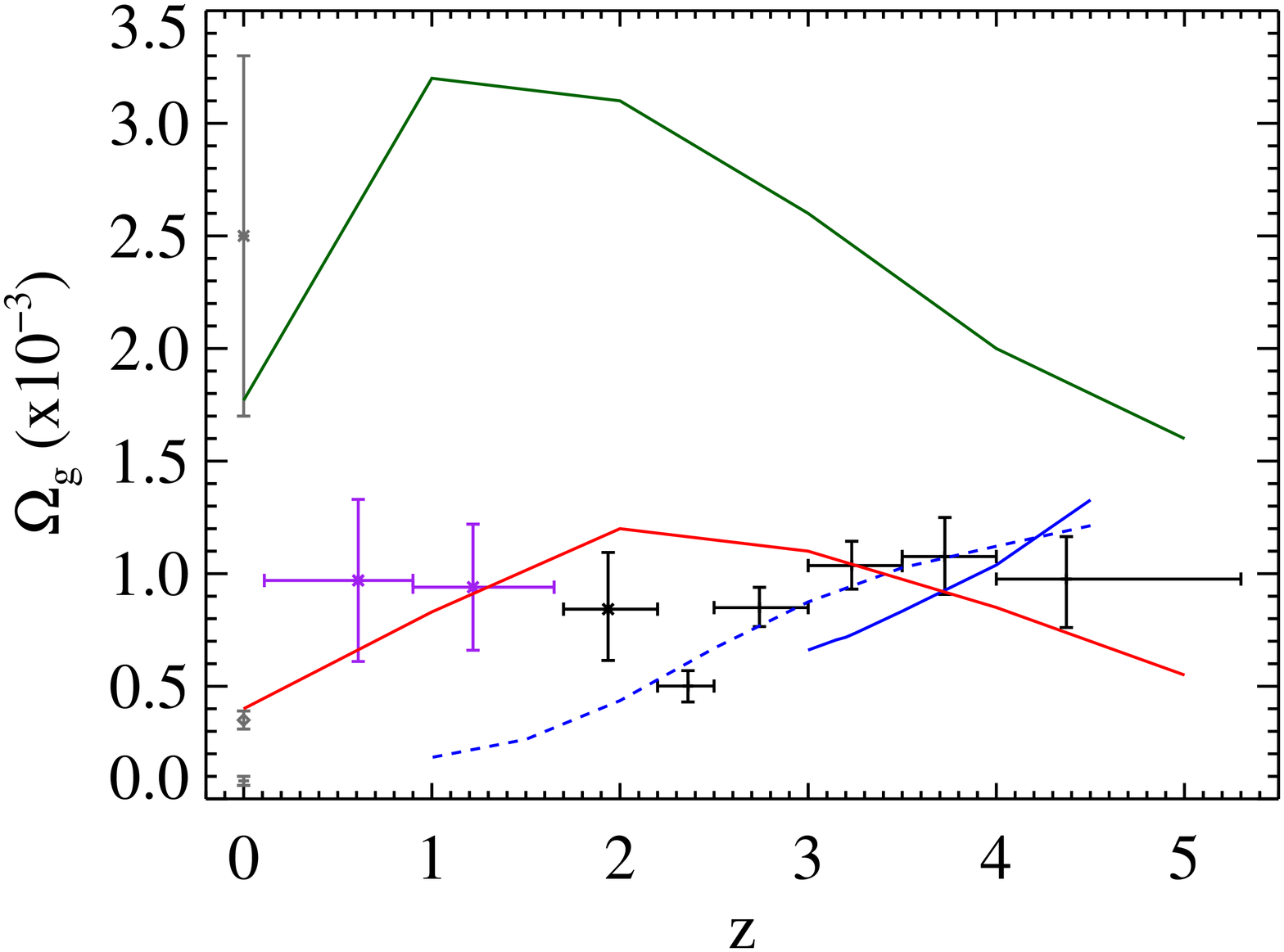}
\caption{
The gas mass density of neutral gas for the damped \lya\ systems
from our analysis ($z>1.5$ with SDSS restricted to $z>2.2$) and
recent results (purple) from Rao, Turnshek, \& Nestor (in prep.).
These observations are 
compared against the theoretical curves of 
\cite{cen03} (EULER; green), \cite{spf01} (SAMS; red),
and \cite{nagamine04a} (SPH; blue -- dotted is the D5 model and
solid is the Q5 model).
The data points at $z=0$ correspond to the stellar mass density
from \cite{cole01} (star), the neutral gas mass density (diamond),
and the mass density of Irr galaxies (+-sign; Fukugita et al.\ 1998).
}
\label{fig:nagomg}
\end{center}
\end{figure*}

Turning our attention to the neutral gas mass density,
we present the damped \lya\ observations in Figure~\ref{fig:nagomg} 
including recent results at $z<1.6$ from Rao, Turnshek, \& Nestor (in prep.)
compared against several $\Lambda$CDM models, and also current estimates
of the mass density of stars (star), neutral gas (diamond), and
Irr galaxies (+ sign) at $z \sim 0$.
It is important to note that the theoretical models of 
Somerville et al.\ and Nagamine et al.\ include contributions to
\omg\ from all quasar absorption line systems (i.e.\ they calculate
\ohi), whereas the observational
measurements are restricted to the damped \lya\ systems.
Therefore, if the Lyman limit systems do contribute significantly
to \ohi, we must increment the observations accordingly.
Alternatively, we recommend that future theoretical analysis be restricted
to sightlines with $\mnhi \geq \Nth$ \citep[e.g.][]{cen03}.

Examining the damped \lya\ observations alone, we note a relatively 
confusing picture.  While the results based primarily on the SDSS-DR3
observations ($z>2.2$) show a well behaved trend with redshift, 
the estimates of \odla\ at $z<2$ are all consistent with one another
with a central value higher than the $z=2.3$ measurement.  
While each individual measurement at $z<2$ is consistent with the SDSS
data point at $z=2.3$, taken together the difference is significant
at $>95\%$ c.l.   In fact, if one
were to ignore the redshift interval at $z=2.3$, the 
observations are consistent
with no evolution in \odla\ from $z=0.1$ to $z=4.5$.
Before reaching such a conclusion, however, we wish to emphasize 
several points:  
(1) the value in the $z = [2.2,2.5]$ interval
is very well determined because it is
based on $\sim 100$ damped \lya\ systems;
(2) the $z \sim 2$ data point is derived from a heterogeneous
sample of observations and is
dominated by a single damped \lya\ system;
(3) the low redshift values are based on the novel yet non-standard
technique of \cite{rao00}.  Their approach has its own set of 
systematic errors which are uniquely different from the damped \lya\
survey described in this paper;
and
(4) we argued in $\S$~\ref{sec:qsomag} that the \odla\ results
may be biased by gravitational lensing.  If this is confirmed, 
the effect should be largest at $z<2$.
These points aside, it is clear that achieving better than 10$\%$
precision on \odla\ at $z<2$ is a critical goal of future damped \lya\
surveys.  
At $z \sim 2$, this will require a large observing campaign with a 
spectrometer efficient down to 3200\AA.  At lower redshift, one will
require a new UV space observatory.

Comparing the models, we note a wide range of predictions.  
The most successful models at $z>2$ are from \cite{nagamine04a}, in particular
their D5~run.  This model reproduces both the shape and normalization
of the observed data.  
In contrast, the Eulerian and SAM models overpredict \odla\ at
all redshifts and at $z<3$ respectively even if one adopts a 
1.5 multiplicative correction due to the Lyman limit systems.
Because the cooling processes and time-scales are comparable in all of the
models \citep[e.g.][]{pearce01}, the differences must be due
to processes which consume or ionize the neutral gas
(e.g.\ star formation, galactic winds, AGN feedback).  
The indication from our observations is that the Eulerian and
SAM models underpredict these processes
at $z \gtrsim 2$ and therefore overestimate \omg.  

Now consider a comparison of the high-$z$ \odla\ values with the mass 
density at $z=0$ of stars \ostr, neutral gas \olwz, and dwarf galaxies 
$\Omega_d$.
The stellar mass density was estimated by \cite{cole01} from an
analysis of the 2dF survey.  The uncertainty in this estimate is
dominated by systematic error related to the assumed initial mass
function.  
\cite{wolfe95} first stressed that the gas mass density of the damped
\lya\ systems is comparable to \ostr.  Adopting the $\Lambda$CDM
cosmology and the current estimate of \ostr, we now find that
\ostr\ exceeds \odla\ by a factor of two to three at $z=3$.
Because star formation is ongoing at all redshifts 
probed by the damped \lya\ systems
\citep[e.g.][]{chen03,wpg03,moller02},
however, it would be wrong to interpret the maximum
\odla\ value as the total gas mass density contributed by the 
damped \lya\ systems.  
It is more accurate to regard damped \lya\ systems as neutral gas
reservoirs in which gas consumed by star formation is replaced
by neutral gas accreted from the IGM.  In this manner, the mass density
of damped \lya\ systems would be less than \ostr\ at any given epoch.
Therefore, it is reasonable to assume
that all of the stars observed today arose from gas originating
in the damped \lya\ systems.  Indeed, this is the generic conclusion
of current cosmological simulations.  These points aside, it is evident
that the damped \lya\ systems contain sufficient gas mass to account 
for all of the stars observed in disks today provided current
estimates ($\Omega_{\rm disk} \approx \Omega_*/3$; 
Fukugita et al.\ 1998). 

Examining Figure~\ref{fig:nagomg}, we note that the difference
between \olwz\ at $z=0$ and \odla\ at $z=2.3$ is $0.15 \pm 0.08$, 
i.e.\ consistent with very little evolution if one ignores 
the results at $z \sim 1$ we present from Rao et al.\ (in prep). 
Is this a remarkable coincidence or is there a physical
explanation (e.g. the
gas accretion rate equaled the star formation rate over the past
10\,Gyr)?  Before addressing this question, consider the determination 
of \olwz.  The value is derived from large area surveys
of 21cm emission-line
observations for \ion{H}{1} `clouds'.  The most recent results
are from the HIPASS survey as analyzed by \cite{zwaan05}.
The analysis proceeds by fitting a functional form (a $\Gamma$-function)
to the \ion{H}{1} mass distribution of all galaxies detected.
The \olwz\ value is simply proportional to 
the first moment of this distribution function.
The key point to emphasize is that the analysis includes all
\ion{H}{1} gas within the beam of HIPASS, i.e., the values are independent
of \ion{H}{1} surface density and are not restricted to the
damped \lya\
threshold.  In the extreme case that the \ion{H}{1} gas is predominantly
distributed in Mestel disks with $\Sigma(R) \propto R^{-1}$, 
the contribution to \olwz\ from damped \lya\ systems is
\odla/\olwz~$= \Sigma_{\rm trunc}/\Sigma_{\rm DLA} = 
N_{\rm trunc}/10^{20.3} \cm{-2}$
where $\Sigma_{\rm trunc}$ and $N_{\rm trunc}$ are the surface density
and \ion{H}{1} column density at the truncation radius of the Mestel disk
and $\Sigma_{\rm DLA}$ is the surface density at 
the damped \lya\ threshold.
One may adopt a truncation radius for the Mestel disk
according to the photoionization
edge set by the intensity of the extragalactic UV background (EUVB)
or observational constraints \citep[e.g.][]{maloney93}.
Adopting a relatively conservative value of $N_{\rm trunc} = 10^{19.5} \cm{-2}$,
we find \odla~=~0.15\olwz.  Of course, this is a somewhat extreme model  
because many gas disks have exponential profiles.  For an exponential
disk, the value of \odla/\olwz\ is sensitive to the ratio of the 
radii corresponding to the damped \lya\ threshold and the 
exponential scale-length.  Taking $R_{DLA}/R_{exp} = 2$, we find
\odla~=~0.5\omg.  These simple estimates imply that a significant
fraction of \olwz\ at $z=0$ could come from Lyman limit systems.

One can also place an empirical constraint on \odla/\omg\ from the
\ion{H}{1} distribution function at $z=0$ as measured by
\cite{ryan03}.  If one (a) adopts their double power-law fit to
\fnhi\ with exponents $\alpha_3 = -1.4$ and 
$\alpha_4 = -2.1$ and $N_d = 10^{20.9} \cm{-2}$, (b) extrapolates
this frequency distribution to $N_{min} = 10^{19} \cm{-2}$, and (c)
integrates to  $N_{max} = 10^{21.6} \cm{-2}$ (the largest \nhi\ value 
measured), we find \odla~=~0.6\omg.
There is significant uncertainty and degeneracy in the fitted parameters
of \fnhi, however, and \odla/\omg\ could be significantly higher
or lower.  Nevertheless, 
it is our present view\footnote{Since submission of our paper, we have
been informed that a better estimate is 
15$\%$ (Zwaan, priv.\ comm.).} that \odla\ at $z=0$ is $\approx 0.5$\omg.
Therefore, we argue that a strict comparison of \odla\ at $z=0$ and $z=2$
could imply a decrease in \odla\ over the past 10\,Gyr.

Given the uncertainties inherent to \ohi\ and the fact that it
has an ambiguous physical meaning,
we now advocate an alternate definition for \omg\ motivated by 
star formation processes.  Define \oneut\ to be the {\it mass
density of gas that is predominantly neutral.}
Under this definition, \oneut\ evolves as follows.
The decrease of the EUVB intensity with decreasing redshift
implies that
lower surface densities will be neutral at lower redshift.
Therefore, it is very possible that 
\oneut\ equals \odla\ at $z=2$ and the HIPASS value at $z=0$.
In this case, the gas reservoir available to star formation may
have remained nearly constant over the past 10\,Gyr (subject, of
course, to the value of \oneut\ from $z=0$ to 2).
We emphasize that this is consistent with the presence of star formation
provided gas replenishment from the IGM is available.
Future progress on these issues will require an accurate determination
of $\flls$ at all redshifts and also an accurate assessment of the
ionization state of the gas as a function of \ion{H}{1} column density.

Finally, compare \odla\ at $z=3$ against the mass density in
dwarf galaxies $\Omega_d$
as derived by \cite{fukugita98}.  The value presented
here is their estimate for the stellar mass density in Irr galaxies
boosted by a factor of 2 to account for the gas mass and by $30\%$ for
$H_0 = 75$\kms\,Mpc$^{-1}$.  We emphasize that the estimate of $\Omega_d$
is difficult to make and may be more uncertain than depicted here.
Nevertheless, it is evident that the mass density of
dwarf galaxies is roughly $10\times$ smaller than \odla\ at $z=3$.
A number of authors have argued that the majority 
of damped \lya\ systems
will evolve into dwarf galaxies today.
It is obvious, however, that the {\it majority of damped \lya\
systems (by mass) cannot evolve into dwarf galaxies}.  And because
damped \lya\ systems with all values of \nhi\ contribute to \odla\
(Figure~\ref{fig:omgcon}), we argue that the damped 
\lya\ systems by number are not the progenitors of dwarf galaxies.
As we have argued previously, it is premature to interpret the
observed abundance patterns as significant evidence for a link
between damped \lya\ systems and dwarf galaxies \citep{pro03}.
Furthermore, the kinematic characteristics of the damped \lya\
systems preclude such an interpretation \citep{pw97,prolmc02}.
Revealing the true nature of the damped \lya\ systems, of course,
will require detailed follow-up studies (high resolution spectroscopy, 
deep imaging, etc.)
of the galaxies identified in this survey.

\acknowledgments

We acknowledge the tremendous effort put forth by the SDSS team to
produce and release the SDSS survey.
JXP would like to acknowledge the visitor's program at Cambridge
where the work on this paper was initiated.
We would like to thank P. McDonald for providing the mock SDSS spectra
for investigating systematic uncertainty in the analysis.
We thank B. Metcalf and A. Maller for helpful
comments and suggestions regarding gravitational lensing and the
damped \lya\ systems.  We thank J. Primack and P. Madau for helpful
discussions.  Additional thanks to M. Murphy for helping to
identify a bug in our search algorithm.
Finally, we wish to thank S. Rao, D. Turnshek, and D. Nestor
for sharing their results in advance of publication.
JXP and SHF (through an REU fellowship) acknowledge 
support through NSF grant AST 03-07408.
JXP and AMW are also partially supported by NSF grant
AST 03-07824.

\end{document}